\documentclass[pdflatex,bst/sn-mathphys]{sn-jnl}

\jyear{2022}%
\usepackage[justification=centering]{caption}
\theoremstyle{thmstyleone}%
\theoremstyle{thmstyletwo}%

\theoremstyle{thmstylethree}%

\newcommand{\ec}{\, ,}

\usepackage{graphicx}

\begin{document}

\title[Ensemble Reservoir Computing for Dynamical Systems]{Ensemble Reservoir Computing for Dynamical Systems: Prediction of Phase-Space Stable Region for Hadron Storage Rings}

\author*[1,2]{\fnm{Maxime} \sur{Casanova}}\email{casanovamaxime@outlook.com}

\author*[1]{\fnm{Barbara} \sur{Dalena}}\email{barbara.dalena@cea.fr}
\equalcont{These authors contributed equally to this work.}

\author[2]{\fnm{Luca} \sur{Bonaventura}}\email{luca.bonaventura@polimi.it}
\equalcont{These authors contributed equally to this work.}

\author[3]{\fnm{Massimo} \sur{Giovannozzi}}\email{massimo.giovannozzi@cern.ch}
\equalcont{These authors contributed equally to this work.}

\affil*[1]{\orgdiv{DRF/Irfu/DACM}, \orgname{CEA Paris Saclay and Paris Saclay University}, \orgaddress{ \city{Gif-Sur-Yvette}, \postcode{91191}, \country{France}}}

\affil[2]{\orgdiv{Dipartimento di Matematica}, \orgname{Politecnico di Milano}, \orgaddress{\street{Via Bonardi 9}, \city{Milano}, \postcode{20132}, \country{Italy}}}

\affil[3]{\orgdiv{Beams Department}, \orgname{CERN}, \orgaddress{\street{Esplanade des Particules 1}, \city{Geneva}, \postcode{1211}, \state{Geneva}, \country{Switzerland}}}

\abstract{We investigate the ability of an ensemble reservoir computing approach to predict the long-term behaviour of the phase-space region in which the motion of charged particles in hadron storage rings is bounded, the so-called dynamic aperture. Currently, the calculation of the phase-space stability region of hadron storage rings is performed through direct computer simulations, which are resource- and time-intensive processes. Echo State Networks (ESN) are a class of recurrent neural networks that are computationally effective, since they avoid backpropagation and require only cross-validation. Furthermore, they have been proven to be universal approximants of dynamical systems. In this paper, we present the performance reached by ESN based on an ensemble approach for the prediction of the phase-space stability region and compare it with analytical scaling laws based on the stability-time estimate of the Nekhoroshev theorem for Hamiltonian systems. We observe that the proposed ESN approach is capable of effectively predicting the time evolution of the extent of the dynamic aperture, improving the predictions by analytical scaling laws, thus providing an efficient surrogate model.}

\maketitle

\section{Introduction}\label{sec1}
The advent of superconducting, high-energy hadron storage rings and colliders elevated non-linear beam dynamics to the forefront of accelerator design and operation. When studying phenomena in the field of single-particle beam dynamics, the concept of dynamic aperture (DA), that is, the extent of the phase space region where bounded motion occurs, has been a key observable to guide the design of several past (see, e.g.\ \cite{visnjic:pac91,Visnjic:1991,brinkmann:epac88,zimmermann:1991jn,zimmermann:1995,luo:pac07-frpms111}), present, e.g.\ the CERN Large Hadron Collider (LHC)~\cite{LHCDR}, and future hadron machines (see e.g.\ ~\cite{appleby:ipac14-tupri013,jing:ipac15-mopmn027,dalena:ipac16-tupmw019,dalena:ipac17-tupva003,cruzalaniz:ipac17-tupva038,TDR,dalena:ipac18-mopmf024,cruzalaniz:ipac18-thpak145}). 

DA prediction involves many challenging aspects, including understanding the mechanisms that determine its behaviour and addressing several computational problems. An important issue is the possibility of modelling the evolution of DA as a function of the number of turns, which has been studied since the end of the $90$s~\cite{dynap1,PhysRevE.57.3432}. Indeed, determining how to describe and efficiently predict the value of the DA might solve some fundamental problems in accelerator physics, linked to performance optimisation of storage rings and colliders. The high computational cost of direct numerical simulations would be significantly reduced if a reliable model for the time evolution of the DA were available. In fact, the numerical simulations required to assess the performance of a circular accelerator cannot cover a time span comparable with operational intervals. For the LHC case, simulations up to $10^6$ turns are at the limit of the CPU-time capabilities, although this represents only about $89$~s of storage time, knowing that a typical fill time is of the order of several hours. Eventually, a model for the evolution of DA over time would also open the possibility of studying observables that are more directly related to machine performance, such as beam losses and lifetime~\cite{PhysRevSTAB.15.024001} and luminosity evolution in colliders~\cite{Giovannozzi:2018wmm,Giovannozzi:2018igq}. 

A successful solution to this problem has been found by building models for DA scaling with time based on fundamental results of dynamical system theory, such as the Nekhoroshev theorem~\cite{Nekhoroshev:1977aa,Bazzani:1990aa,Turchetti:1990aa}. In fact, models with two or three parameters can be derived that can be fitted to numerical data that represent the evolution of the DA and used to predict the DA value for times beyond the current computational capabilities~\cite{ScalingLaws}.

In the last decade, the use of neural networks has increased significantly in a large number of diverse research areas, and this observation has suggested their application to the prediction of the evolution of DA. For example, neural networks are used for speech recognition~\cite{bib2} or to forecast wind power~\cite{ensemble_val}. Among neural network techniques, the most common architectures are feedforward~\cite{fnn}, convolutional~\cite{cnn}, and recurrent~\cite{rnn} neural networks. Feedforward neural networks are made up of neurons connected to other neurons, only. They provide only input-output relationships and can approximate very large classes of functions. On the other hand, recurrent neural networks are made up of neurons connected to themselves and other neurons. They preserve an internal state that is a non-linear transformation of the input signal and can therefore be considered as dynamical systems.

Echo State Networks (ESN) are one of the classes of recurrent neural networks that use the reservoir computing approach~\cite{LUKOSEVICIUS2009127}. This approach has the main advantage of significantly reducing the computational time required by the training process, which is performed to find the optimal parameters (called weights) of a neural network. In fact, the peculiarity of the ESN is that training is performed, usually using linear regression~\cite{esn_linear}, to calculate the weights used to project the reservoir state onto the output state. Therefore, no backpropagation is needed. Backpropagation~\cite{back} refers to the numerical procedure, usually based on the stochastic gradient method, used for the training of feedforward networks, which is responsible for a large share of its computational cost. ESN have also been proven to be universal approximants of dynamical systems~\cite{ESNuniversal}. Thus, ESN seem to be natural candidates for performing the prediction of DA for a large number of turns, and hence challenge the performance of the deterministic models developed so far.
 
This paper is organised as follows: In Section~\ref{sec2}, we introduce the concept of DA and the approach used to provide numerical estimates of its value. Analytical scaling laws, based on the Nekhoroshev theorem and used to predict the time evolution of DA, are also presented. Section~\ref{sec3} introduces the continuous-time leaky ESN framework that is used for the prediction of DA. The Echo State Property (ESP), and a sufficient condition that can be applied in practice to satisfy it, are discussed in the Appendix~\ref{AP:ESP}. Section~\ref{sec4} describes the ensemble procedure used in the cross-validation of the ESN and in the prediction of DA. The results are presented and discussed in Section~\ref{sec5}, while conclusions are drawn in Section~\ref{conc}.
\section{Dynamic Aperture}\label{sec2}
\subsection{Generalities} \label{subsec1}
We consider a Hamiltonian system in $\mathbb{R}^{2n}$, with a stable fixed point at the origin, whose dynamics is generated by a polynomial map $\mathcal{M}$, and 
 such that the linear part of $\mathcal{M}$ is described by the direct product of rotations. Under these conditions, the DA of the system under consideration is the extent of the region of phase space in which bounded motion occurs. 

Following~\cite{bib1} and restricting the analysis to the case of Hamiltonian systems in $\mathbb{R}^4$, which are relevant for accelerator physics, we consider the phase space volume of the initial conditions that are bounded after $N$ iterations, namely
\begin{equation}
\int \int \int \int \chi(x_1,p_{x_1},x_2,p_{x_2}) \; dx_1 \, dp_{x_1}  \, dx_2 \, dp_{x_2} \, ,
\label{eqn8o}
\end{equation}
where $\chi(x_1,p_{x_1},x_2,p_{x_2})$ is the characteristic function defined as equal to one if the orbit starting at $(x_1,p_{x_1},x_2,p_{x_2})$ is bounded and zero if it is not. 

To exclude the disconnected part of the stability domain in the integral~\eqref{eqn8o}, we have to choose a suitable coordinate transformation. As linear motion is given by the direct product of constant rotations, the natural choice is to use the polar variables $(r_i,\vartheta_i)$, where $r_1$ and $r_2$ are the linear invariants of dynamics. The non-linear part of the equations of motion adds a coupling between the two planes, the perturbative parameter being the distance from the origin. Therefore, it is natural to replace $r_1$ and $r_2$ by the polar variables $r \cos \alpha$ and $r \sin \alpha$, respectively:
\begin{equation}
\left \{ \begin{array}{lcll}
  x_1 &=& r \cos \alpha  \cos \vartheta_1  &           \\
p_{x_1} &=& r \cos \alpha  \sin \vartheta_1  &      \qquad \qquad r \in [0,+\infty[             \\
  & &  &           \qquad \qquad \alpha \in [0,\pi/2]      \\
  x_2 &=& r \sin \alpha  \cos \vartheta_2  &           \qquad \qquad \theta_i \in [0,2\pi[ \qquad i=1,2      \\
p_{x_2} &=& r \sin \alpha  \sin \vartheta_2    \, .  &        
\end{array} \right .
\label{eq00o}
\end{equation}
Substituting in Eq.~\eqref{eqn8o} we obtain
\begin{equation}
\int_0^{2\pi} \int_0^{2\pi} \int_0^{\pi/2}\int_0^\infty 
\; \chi(r, \alpha, \vartheta_1, \vartheta_2) \, r^3 \sin \alpha \cos \alpha \; d \Omega_4 \, , 
\end{equation}
where $d\Omega_4$ represents the volume element
\begin{equation}
d\Omega_4 = dr \, d\alpha \, d\vartheta_1 \, d\vartheta_2 \, .
\end{equation}
Having fixed $\alpha$ and $\boldsymbol{\vartheta}=(\vartheta_1,\vartheta_2)$, let $r(\alpha, \boldsymbol{\vartheta},N)$ be the last value of $r$ whose orbit is bounded after $N$ iterations. Then, the volume of a connected domain in which the motion is bounded is given by 
\begin{equation}
A_{\alpha,\boldsymbol{\vartheta},N} = \frac{1}{8} \, \int_0^{2\pi} \int_0^{2\pi} \int_0^{\pi/2} [r(\alpha,\boldsymbol{\vartheta},N)]^4 \sin 2 \alpha \; d\Omega_3 \, ,
\label{eqn3o}
\end{equation}
where
\begin{equation}
d\Omega_3 = d\alpha \, d\vartheta_1 \, d\vartheta_2 \, .
\end{equation}
In this way, we exclude stable islands that are not connected to the main stable domain. Note that, in principle, this method might also lead to excluding connected parts. We then define the DA as the radius of the hypersphere that has the same volume as the stability domain
\begin{equation}
r_{\alpha,\boldsymbol{\vartheta},N} = \left( \frac{2 A_{\alpha,\boldsymbol{\vartheta},N} }{\pi^2} \right)^{1/4} \, .
\label{radius4D}
\end{equation}

When Eq.~\eqref{eqn3o} is implemented in a computer code, one considers $K$ steps in the angle $\alpha$ and $L$ steps in the angles $\vartheta_i$, and the dynamic aperture reads
\begin{equation}
r_{\alpha,\boldsymbol{\vartheta},N}  = \left[ \frac{\pi}{2 \,K L^2} \sum_{k=1}^{K} \sum_{l_1,l_2=1}^L 
[r(\alpha_k,\boldsymbol{\vartheta}_\mathbf{\ell},N)]^4 \sin 2 \alpha_k \right]^{1/4} \, , 
\nonumber
\end{equation}
where $\mathbf{\ell}=(l_1,l_2)$. 

The numerical error is given by the discretization in angles $\vartheta_i$, $\alpha$, and radius $r$, which gives a relative error proportional to $L^{-1}$, $K^{-1}$, and $J^{-1}$, respectively. This numerical error can be optimised by choosing integration steps that produce comparable errors, i.e. $J \propto K \propto L$. In this way, neglecting the constants in front of the error estimates, one can obtain a relative error of $1/(4J)$ by evaluating the $J^4$ orbits, i.e. $N J^4$ iterates.
The fourth power in the number of orbits comes from the dimensionality of phase space and makes a precise estimate of the dynamic aperture very CPU time consuming.

It is possible to reduce the size of the scanning procedure, and hence the CPU time needed, by setting the angles $\boldsymbol{\theta}$ to a constant value, e.g. zero, thus performing only a 2D scan over $r$ and $\alpha$. This is what is generally done in SixTrack simulations~\cite{sixtrack,DeMaria:2711546}. In this case, the transformation \eqref{eq00o} reads
\begin{equation}
\left \{ \begin{array}{lcll}
  x_1 &=& r \cos \alpha   &           \\
p_{x_1} &=& 0  &           \qquad \qquad r \in [0,+\infty[                     \\
  x_2 &=& r \sin \alpha  & \qquad \qquad \alpha \in [0,\pi/2] \\          
p_{x_2} &=& 0  \, ,&             
\end{array} \right .
\end{equation}
and the original integral is transformed to 
\begin{equation}
\int_0^{\pi/2}\int_0^\infty \; r \; d r \, d\alpha \, . 
\end{equation}
Having fixed $\alpha$, let $r(\alpha,N)$ be the last value of $r$ whose orbit is bounded after $N$ iterations. Then, the volume of a connected stability domain is given by
\begin{equation}
A_{\alpha,N} =  \frac{1}{2} \int_0^{\pi/2} [r(\alpha,N)]^2 \; d\alpha \, .
\label{eqn3oo}
\end{equation}
We define the dynamic aperture as the radius of the sphere that has the same volume as the stability domain\footnote{Note that the region providing the stability domain is confined to a surface that is $1/4$ of a circle and this has been considered in Eq.~\eqref{eqnpippo}.}
\begin{equation}
r_{\alpha,N} = \left(\frac{4 A_{\alpha,N} }{\pi} \right)^{1/2} \, .
\label{eqnpippo}
\end{equation}

When Eq.~\eqref{eqn3oo} is implemented in a computer code, one considers $K$ steps in the angle $\alpha$, and the dynamic aperture reads
\begin{equation}
r_{\alpha,N}  = \left[ \frac{1}{K}
\sum_{k=1}^{K} [r(\alpha_k,N)]^2 \right]^{1/2} \, , 
\end{equation}
so that the numerical error is given by discretising the angle $\alpha$ and the radius $r$, which yields a relative error proportional to $K^{-1}$ and $J^{-1}$, respectively. In this case, the integration steps should also be selected to produce comparable errors, i.e. $J \propto K$. In this way, neglecting the constants which are in front of the error estimates, one can obtain a relative error of $1/(2J)$ by evaluating $J^2$ orbits, i.e. $N J^2$ iterates\footnote{The factor $2$ in the error estimate is due to the dimensionality of the phase space}. Note that Eq.~\eqref{eqn3oo} can be evaluated using higher-order numerical integration rules as implemented in the post-processing tools linked with SixTrack~\cite{DeMaria:2711546}.

It is worth noting that, in some applications, the simplified formula 
\begin{equation}
r_{\alpha,N}  = \frac{1}{K} \sum_{k=1}^{K} [r(\alpha_k,N)] \, ,
\end{equation}
 which corresponds to computing the average of $r(\alpha_k,N)$ over the angle $\alpha_k$, could be used~\cite{PhysRevE.57.3432}. 
\subsection{DA Scaling Law}\label{subsec2}
All the definitions of DA estimates presented in the previous section are functions of $N$, the turn number used to estimate the orbit stability from the results of numerical simulations. It is evident that the definition of DA itself implies that it is a non-increasing function of $N$. The key point is whether it is possible to find the functional form of this time dependence, and several studies have shown that this is indeed the case~\cite{PhysRevE.57.3432,ScalingLaws}. In fact, such a functional form can be built by considering the estimate of the stability time provided by the Nekhoroshev theorem~\cite{Nekhoroshev:1977aa,Bazzani:1990aa,Turchetti:1990aa}, which is a key and very general theorem in the theory of Hamiltonian dynamical systems. 

The first models were described in~\cite{PhysRevE.57.3432} and then reviewed in~\cite{ScalingLaws} and the two that we retained after the review read
\begin{equation}
\begin{split}
\textbf{Model 2} \qquad \Rightarrow \qquad D(N) & =\rho_\ast \left ( \frac{\kappa}{2 \text{e}} \right )^\kappa \, \frac{1}{ \ln^\kappa \frac{N}{N_0}} \ec
\label{model2.1_1}
\end{split}
\end{equation}
where the free parameters are $\rho_\ast, \kappa, N_0$, but it is customary to set $N_0=1$, and 
\begin{equation}
\begin{split}
& \;\, \textbf{Model 4} \qquad \Rightarrow \qquad D(N) = \rho_\ast \times \\
& \times \displaystyle{\frac{1}{\left[-2 \, \text{e} \, \lambda \,\mathcal{W}_{-1}\!\!\:\!\left(
-\frac{1}{2 \, \text{e} \, \lambda}\left( \frac{\rho_\ast}{6} \right)^{1/\kappa} \, \left( \frac{8}{7} N \right)^{-1/(\lambda \, \kappa)} \right)
		\right]^{\kappa}}} \ec \phantom{\times}
\end{split}
\label{eq: new model exact}
\end{equation}
where the free parameters are $\rho_\ast$, $\kappa$, and possibly $\lambda$, unless it is fixed to the value of $1/2$ according to the analytic Nekhoroshev estimate. $\mathcal{W}_{-1}$ stands for the negative branch of the Lambert-$\mathcal{W}$ function, a multi-valued special function (see, e.g.\ \cite{corless:1996} for a review of the properties and applications of the Lambert function). Note that $D(N)$ stands for $r_{\alpha,\boldsymbol{\vartheta},N}$ or $r_{\alpha,N}$, depending on the numerical approach used to estimate the DA. The nomenclature of the models presented in Eqs.~\eqref{model2.1_1} and~\eqref{eq: new model exact} reflects the historical development of these models and the nomenclature used in~\cite{ScalingLaws}. 

\begin{figure}[!htb]
\centering
\includegraphics[width=5.3cm]{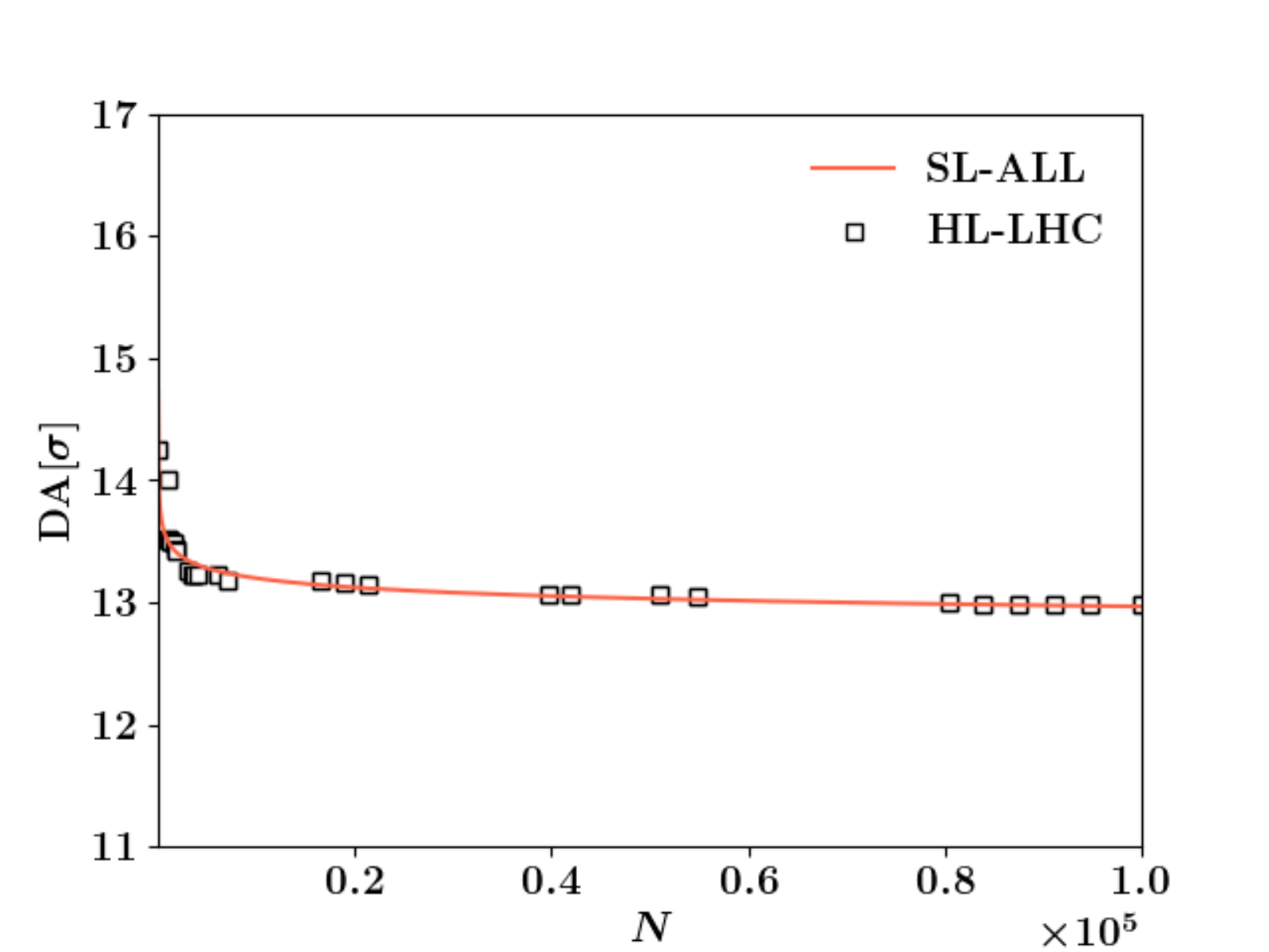}
\centering
\caption{Example of DA numerical computation for a realistic model of the HL-LHC with the corresponding fitted scaling law. The excellent agreement between the numerical data and the fit model is clearly visible.}
\label{fig:hl}
\end{figure}

An example of the numerical calculation of the DA for a realistic model of the luminosity upgrade of the CERN LHC, HL-LHC~\cite{TDR}, and the corresponding fitted scaling law using all available DA data are shown in Fig.~\ref{fig:hl}, where the excellent agreement between the numerical data and the fit model is clearly visible. We denote by SL-ALL the fitting of \textbf{Model 2} using all available DA data.

Note that in the rest of the paper \textbf{Model 2} is the only scaling law model used. 
%
\subsection{DA data organisation}
In this section, we present the data sets used to test the predictive model introduced in Section \ref{sec4}. The first data set is obtained from a realistic model of the HL-LHC, whereas the second one is obtained from the 4D H\'enon map.

\subsubsection{The HL-LHC case}

\begin{figure}[!htb]
\minipage{0.5\textwidth}
\centering
  \includegraphics[width=5.3cm]{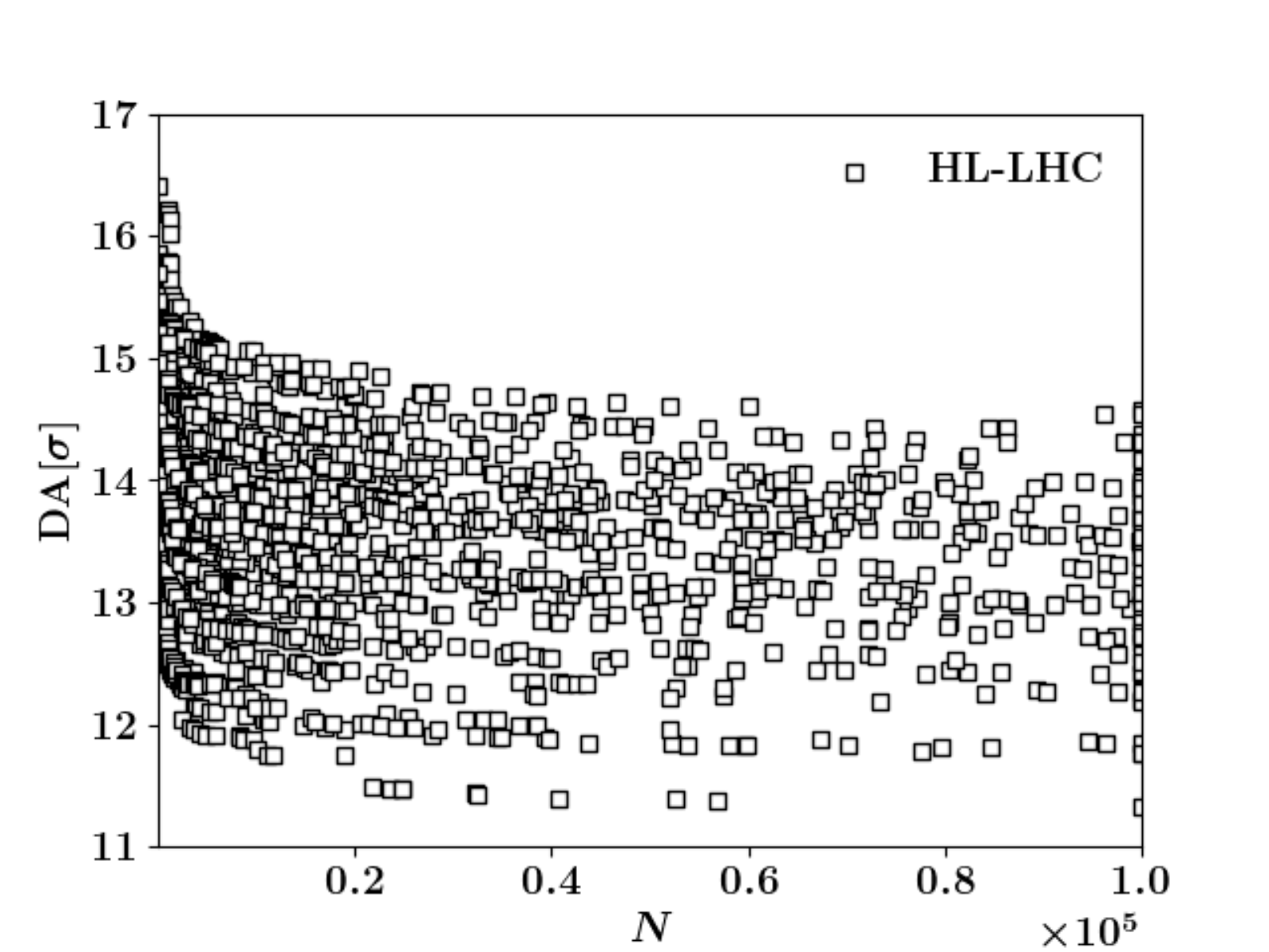}
\endminipage\hfill
\minipage{0.5\textwidth}
\centering
  \includegraphics[width=5.3cm]{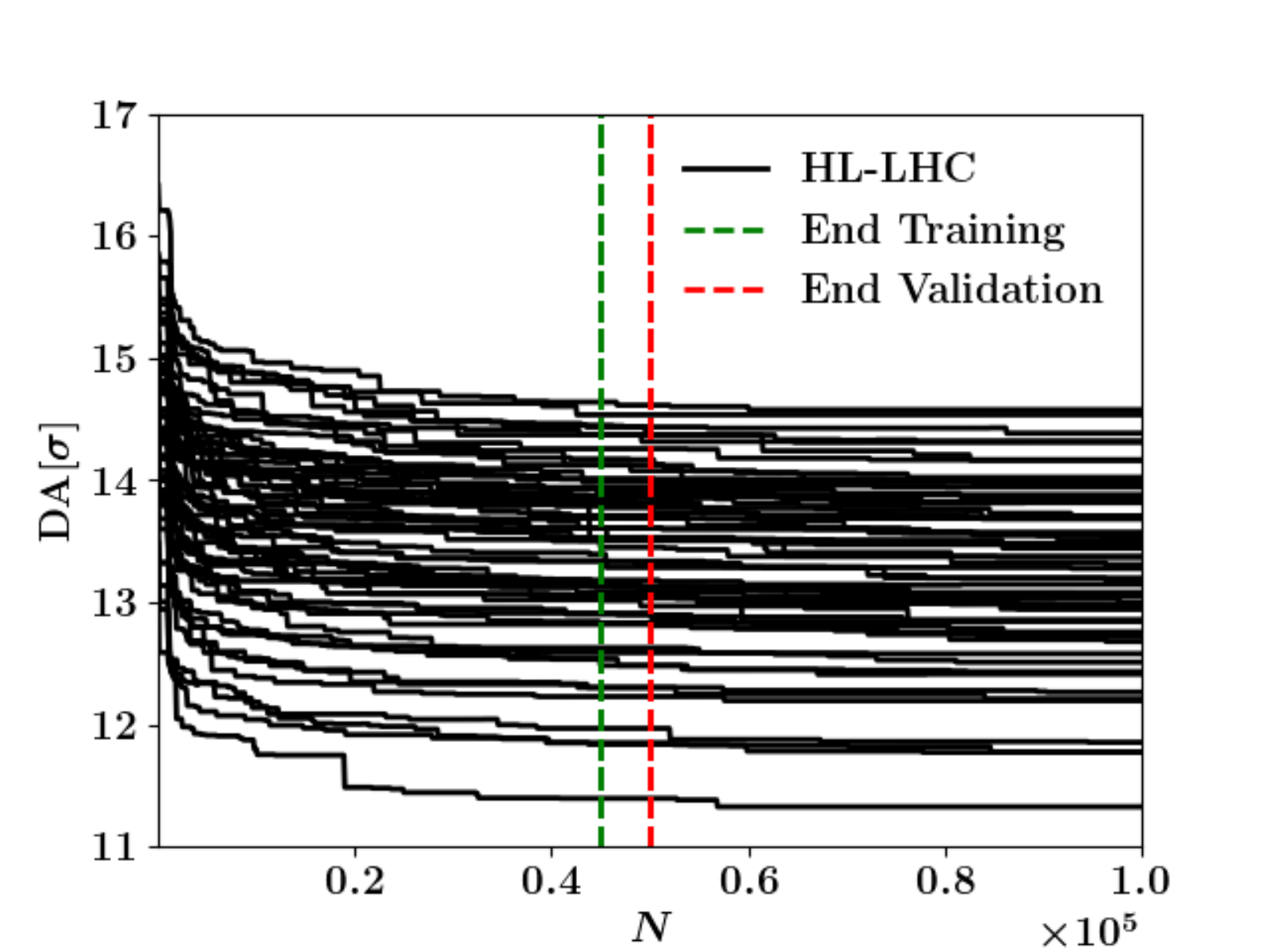}
  \endminipage\hfill
  \caption{Left: Evolution of DA as a function of time for the 60 realisations of the HL-LHC magnetic lattice. Right: Splitting of the HL-LHC data set into training, validation, and test sets.}
  \label{fig:hldata}
\end{figure}

The HL-LHC data set, presented in Fig.~\ref{fig:hldata}, is composed of 60 realisations (also called seeds due to the underlying random generator used for the generation of the realisations) of the magnetic field errors of the magnetic lattice of the HL-LHC, for the collision optics with $\beta^{*}$=15 cm and proton energy of $7$~TeV. The 60 realisations are supposed to accurately represent the actual lattice of the HL-LHC; for this reason, the DA computation is customarily performed using the complete set of realisations to provide an accurate estimate of the DA of the actual accelerator. Magnetic field errors are assigned to all magnets that make up the ring. Initial conditions (also called particles) are distributed in physical space to probe the orbit stability and thus determine the DA. Different amplitudes and angles in the $x-y$ plane are used to sample the phase space. In the cases considered here, $11$ angles, uniformly distributed in the interval $]0, \pi/2[$, are used, while the amplitudes are uniformly distributed in the interval $]0,28 \sigma[$, with $30$ initial conditions defined in each 2$\sigma$ interval. Note that $30$ particles are evenly distributed in each amplitude interval of $2\sigma$, and $\sigma$ represents the root mean square (rms) beam size, which is used as a natural unit in these studies. All initial conditions are tracked for $10^5$ turns. The numerical estimates of DA as a function of $N$ are calculated according to Eq.~\eqref{eqn3oo} and are shown in Fig.~\ref{fig:hldata} (left). 

We build piecewise constant functions so that each DA estimate now contains $10^3$ data points, with the aim of obtaining DA estimates in constant time steps. These $10^3$ data points are then divided into \textit{training set}, \textit{validation set}, and \textit{test set}. The first $k_\mathrm{train} = 450$ data are used for training, the next $k_\mathrm{val} =  50$ data for validation, and the remaining $k_\mathrm{test} = 500$ data for testing. Note that the end of the training and validation sets corresponds to $N = 5.10^4$, and the end of the testing to $N = 10^5$ turns. A graph of the 60 piecewise constant functions split into \textit{training set}, \textit{validation set} and \textit{test set} is shown in Fig.~\ref{fig:hldata} (right). Note that each of the 60 realisations corresponds to a different DA on which we will train, validate, and test our ESN model.

\subsubsection{The 4D H\'enon map case}

The 4D H\'enon map is a well-known dynamical system that displays a rich dynamical behaviour as presented in, e.g.~\cite{Bazzani:262179}. The model used to generate DA estimates is defined as:
\begin{center}
\begin{equation}
    \begin{pmatrix}
x_{n+1} \\
p_{x, n+1} \\
y_{n+1}\\
p_{y, n+1}\\
\end{pmatrix}
=
\widetilde{R} 
    \begin{pmatrix}
x_{n} \\
p_{x, n}  + x_{n}^2 - y_{n}^2 + \mu \left ( x_{n}^3 - 3 y_{n}^2 x_{n} \right)\\
y_{n}\\
p_{y, n} - 2 x_{n} y_{n} + \mu \left( y_{n}^3 - 3 x_{n}^2 y_{n} \right)\
\end{pmatrix}
\end{equation}
\end{center}
where the subscript $n$ denotes the discrete time and $\widetilde{R}$ 
 is a $4\times 4$ matrix given by the direct product of two $2\times 2$ rotation matrices R:
\begin{center}
\begin{equation}
\widetilde{R} 
  =
    \begin{pmatrix}
R(\omega_{x, n}) & 0\\
0 & R(\omega_{y, n})
\end{pmatrix} \, ,
\end{equation}
\end{center}
where the linear frequencies vary with the discrete time $n$ according to
\begin{align}
\omega_{x, n} & = \omega_{x, 0} \left (1+\varepsilon \sum_{k=1}^{m} \varepsilon_k cos(\Omega_k n) \right)
\label{eq:modx}\\
\omega_{y, n} & = \omega_{y, 0} \left (1+\varepsilon \sum_{k=1}^{m} \varepsilon_k cos(\Omega_k n) \right) \, , 
\label{eq:mody}
\end{align}
where $\varepsilon$ denotes the amplitude of the frequency modulation and $\varepsilon_k$ and $\Omega_k$ are fixed parameters, which are taken from previous studies~\cite{ScalingLaws}\footnote{Note that all $\varepsilon_k$ are of order $10^{-4}$. Therefore, even if $\varepsilon$ is large, the effective modulation of the frequencies shown in Eqs.~\eqref{eq:modx} and~\eqref{eq:mody} is very small.}.

\begin{figure}[!htb]
\centering
\includegraphics[width=5.3cm]{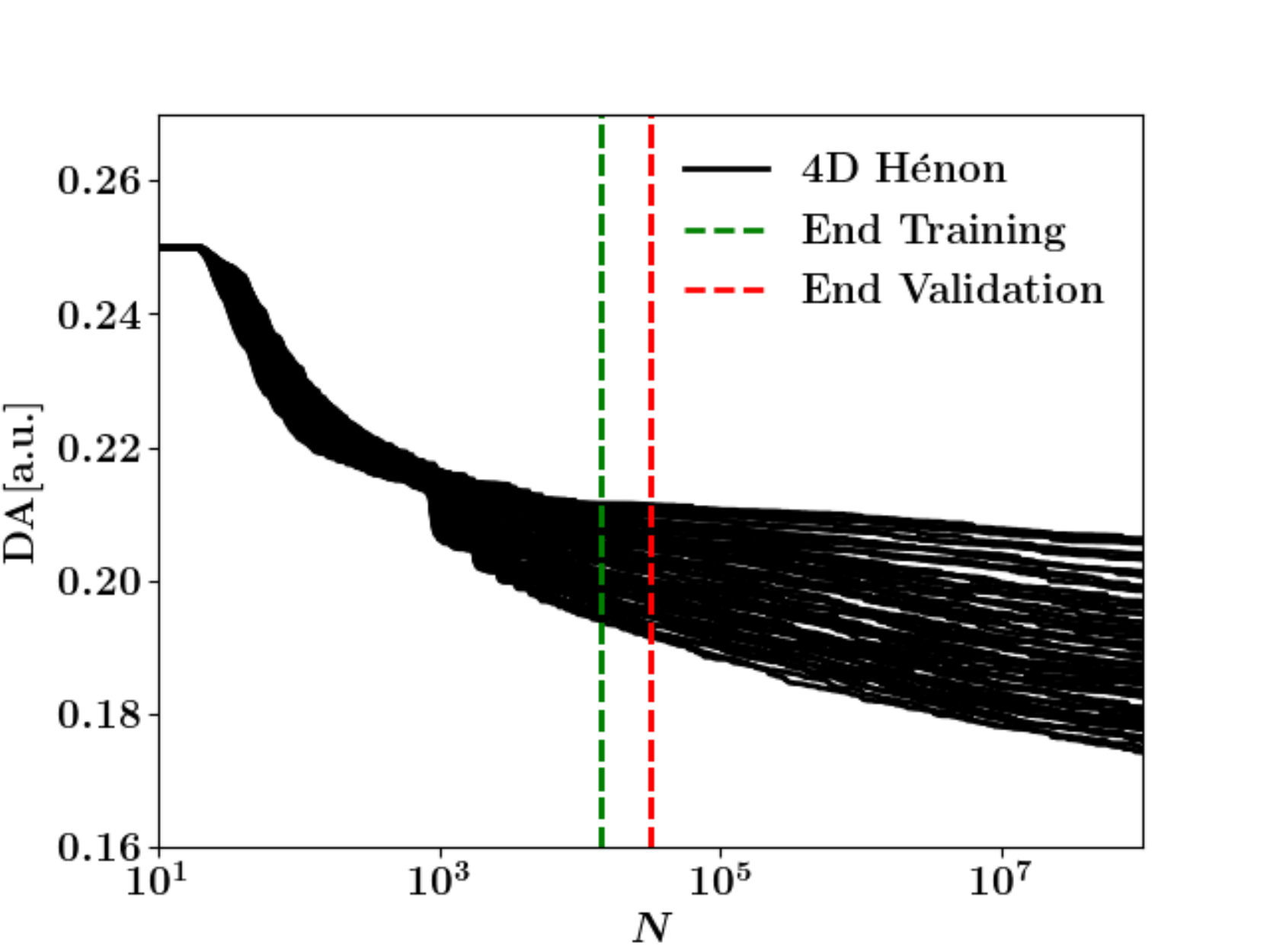}
\centering
\caption{Splitting of the 4D H\'enon map data set into training, validation, and test sets. The sudden drop in DA visible for $N \approx 10^3$ occurs when $\varepsilon > 15$.}
\label{fig:henonsplit}
\end{figure}

The 4D H\'enon map is a simplified model of a circular accelerator. In particular, it describes the effects of a sextupole and octupole magnet on the transverse particle motion through the quadratic, due to the sextupole, and cubic, due to the octupole, non-linear terms. Being a simplified accelerator model, it allows one to track particles up to a much larger number of turns, and for more amplitudes and angles, namely $100$ amplitudes and angles uniformly distributed in the interval $]0, 0.25[$ and $]0, \pi/2[$ respectively. The 4D H\'enon map data set is composed of 60 cases, for 20 different values of $\varepsilon$ uniformly distributed in the interval $[0, 20[$ and $\mu \in \{-0.2, 0, 0.2\}$, covering up to $10^8$ turns. Similarly to the HL-LHC data set, we build piecewise-constant functions so that each case yields 1000 data points. The first $k_\mathrm{train} =  450$ data are used for training, the next $k_\mathrm{val} =  50$ data for validation, and the last $k_\mathrm{test} =  500$ data for testing.  Note that we used the same number of training, validation, and test data for the HL-LHC case.
The 60 piecewise constant functions divided into \textit{training set}, \textit{validation set}, and \textit{test set} are shown in Fig.~\ref{fig:henonsplit}. 

Note that because of the larger number of amplitudes and angles considered, the DA data are smoother than those of the HL-LHC case. Furthermore, each of the 60 cases generated in this data set corresponds to a different dynamics for which we will train, validate, and test our ESN model.

\section{Echo State Networks}\label{sec3}

In this section, we present some general concepts about ESN. More specifically, we introduce the mathematical framework of continuous-time leaky ESN applied to supervised learning tasks. 

\subsection{Shallow ESN}

Shallow ESN are a class of Recurrent Neural Networks using the Reservoir Computing approach~\cite{LUKOSEVICIUS2009127}. In this type of neural network, the data input is fed into a single, random, and non-trainable network, called the reservoir. The reservoir is eventually connected by trainable weights to the ESN output. The use of ESN for time series prediction has become widespread due to its inexpensive training process and its remarkable performance in the modelling of dynamical systems~\cite{esn_time_serie}.

Contrary to feedforward neural networks, ESN do not suffer from vanishing or divergent gradients (caused by the fact that the parameters of neural networks remain almost constant or lead to numerical instabilities), which induces poor performance of the training algorithm~\cite{exploding}.

ESN can be defined for discrete- or continuous-time systems. The reservoir dynamics can be defined with or without the leaking rate parameter, which can be considered as the speed of the reservoir update dynamics. We introduce the definition of a shallow leaky ESN in continuous time as in~\cite{JAEGER2007335}. We consider the case of networks with continuous-time $t$, $K$ inputs, $N_\mathrm{r}$ reservoir neurons, and $M$ outputs.  Note that we will use small letters to indicate vectors and capital letters to indicate matrices. We define by $u = u(t) \in \mathbb{R}^K$ the input data and $x^{\mathrm{train}}$ = $x^{\mathrm{train}}(t) \in \mathbb{R}^M$ the training data that we want to learn with the ESN model. The ESN output is denoted by $x^{\mathrm{out}} =x^{\mathrm{out}}(t) \in \mathbb{R}^M$,  while the internal reservoir activation state is given by $x = x(t) \in \mathbb{R}^{N_\mathrm{r}}$. Furthermore, we define the input weight matrix $W^{\text{in}} \in {\cal M}_{N_\mathrm{r}\times K}(\mathbb{R})$, the reservoir weight matrix $W \in {\cal M}_{N_\mathrm{r}\times N_\mathrm{r}}(\mathbb{R})$, and the output weight matrix $W^{\text{out}} \in  {\cal M}_{M\times (N_\mathrm{r}+K)}(\mathbb{R})$. 
The discretised (by the Euler method) time dynamics of a leaky ESN is given by:
\begin{align}
    x_{k} & = F(x_{k-1},u_{k}) = \left(1-a\Delta t\right)x_{k-1} + \Delta t f(W^{\mathrm{in}}u_{k} + Wx_{k-1})\label{eq:discrESN}
    \\
    x_{k}^{\mathrm{out}} & = g(W^{\mathrm{out}}[x_{k};u_k]) \label{eq:discrESN2}
\end{align}
where $\Delta t$ = $\delta/c$ with $\delta$ the size of the Euler discretization step and $c$ a global time constant, $a$ the leaking rate, $f$ a sigmoid function, $g$ the output activation function, $[.;.]$ denotes vector concatenation, $x_{k}$ the update of the reservoir activation state at discrete time $k$ and $x^{\mathrm{out}}_{k}$ the ESN output at the same time $k$.

In the case of a linear readout, i.e. when $g$ is the identity function, we can rewrite Eq.~\eqref{eq:discrESN2} in matrix notation as:
\begin{align}\label{eq:trainESNout}
    &X^{\mathrm{out}} = W^{\mathrm{out}}X
\end{align}
where $X^{\mathrm{out}} \in {\cal M}_{M\times (k_{\mathrm{train}}-BI)}(\mathbb{R})$ contains the $M$ ESN outputs $x^{\mathrm{out}}$ at every time step $k=BI,\ldots,k_{\mathrm{train}}$ and where $X \in {\cal M}_{(N_\mathrm{r}+K)\times (k_{\mathrm{train}}-BI)}(\mathbb{R})$  contains the concatenation of the input $u$ and the activation state of the reservoir $x$ at every $k=BI,\ldots,k_{\mathrm{train}}$, namely
\begin{align}\label{eq:trainESN}
    & X = \begin{pmatrix}
u_{BI} & \ldots & u_{k_{\mathrm{train}}}\\
x_{BI} & \ldots & x_{k_{\mathrm{train}}+1}
\end{pmatrix} \, , 
\end{align}
where $BI$ denotes the \textit{Burn-In} data, i.e. the number of input data we want to discard at the beginning of the training phase. 

The optimal output weight matrix $W^{\mathrm{out}}$  can be found by solving the following minimisation problem:
\begin{equation}\label{eq:minimzation}
\begin{split}
        W^{\text{out}} & = \operatorname*{argmin}  J(W^{\mathrm{out}}) \\
        & = \operatorname*{argmin} \frac{1}{M} \sum_{i=1}^{M}\Big(\sum_{k=BI}^{T}(x_{ik}^{\mathrm{out}} - x_{ik}^{\mathrm{train}})^2 + \beta\|w_i^{\mathrm{out}}\|^2\Big) \, , 
\end{split}
\end{equation}
where $J$ denotes the cost function we want to minimise and $\|w_i^{\mathrm{out}}\|$ is the Euclidean norm of the $i$th row of $W^{\mathrm{out}}$.

The solution of the minimisation problem stated in Eq.~\eqref{eq:minimzation}  can be found efficiently using linear regression with Tikhonov (Ridge) regularisation~\cite{regul}:
\begin{equation}\label{eq:ridge}
    W^{\mathrm{out}} = X^{\mathrm{train}}X^{T}(XX^{T}+\beta I)^{-1}
\end{equation}
where the superscript $T$ denotes the transpose, $I \in {\cal M}_{(N_\mathrm{r}+K)\times (N_\mathrm{r}+K)}(\mathbb{R})$ is the identity matrix, and $X^{\text{train}} \in {\cal M}_{M\times (k_{\text{train}}-BI)}(\mathbb{R})$ is the training data matrix, which contains the $M$ training data $x^{\text{train}}$ at time step $k = BI, \ldots, k_\mathrm{train}$.

\begin{figure}[!htb]
\centering
\includegraphics[width=11cm]{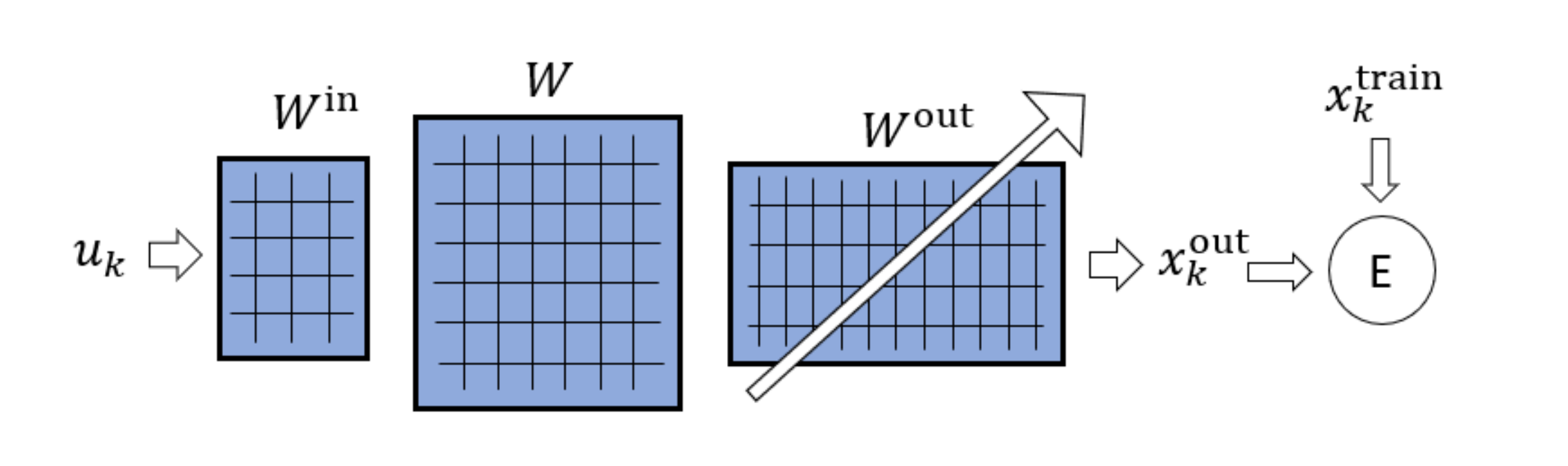}
\centering
\caption{Sketch of the training procedure for a shallow leaky ESN. The size of the matrices has been arbitrarily selected. $E$ denotes the square of the Euclidean norm error between the ESN output $x_{k}^{\mathrm{out}}$ and the training data $x_{k}^{\mathrm{train}}$, $k = BI ,\ldots , k_{\mathrm{train}}$.}
\label{fig:ESNtrain}
\end{figure}

The learning phase is carried out on the so-called \textit{training set}, which contains the $k_{\mathrm{train}}$ training data $x^{\mathrm{train}}$. A sketch of the training phase of the ESN is provided in Fig.~\ref{fig:ESNtrain}. 

After training, the ESN hyperparameters, defined in Section \ref{sec4}, are tuned using $k_{\mathrm{val}}$ validation data. Finally, the ESN is tested using the $k_{\mathrm{test}}$ data to check the ability of the ESN to predict new data. The validation and test procedures are detailed in Section~\ref{sec4}. As stated in Eq.~\eqref{eq:minimzation}, only the output weight matrix $W^{\mathrm{out}}$ is trained, while the input and reservoir matrices $W^{\mathrm{in}}$ and $W$ are randomly generated, as explained in detail in Section~\ref{sec4}.

\subsection{Deep ESN}

\begin{figure}[!htb]
\centering
\includegraphics[width=11cm]{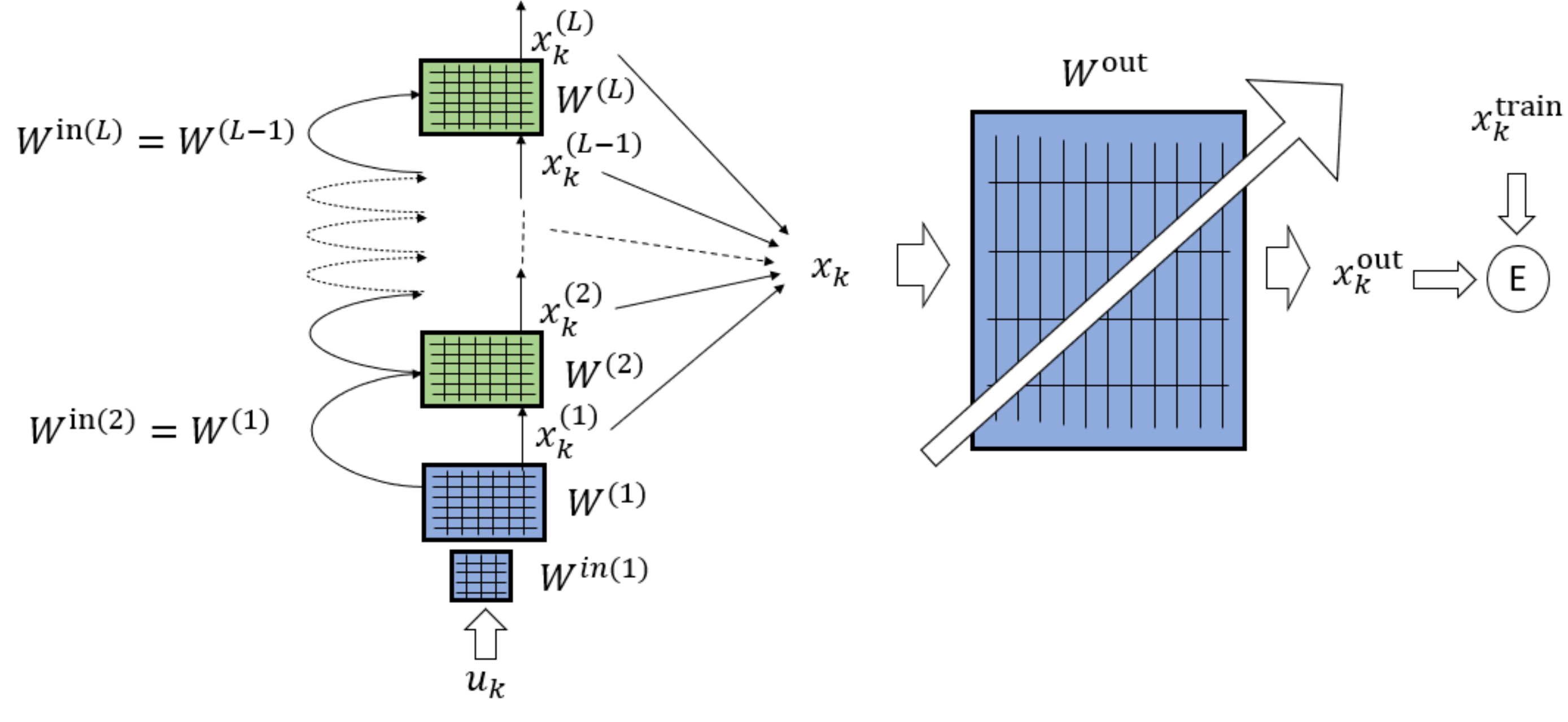}
\centering
\caption{Sketch of the training procedure for deep ESN with $L$ reservoirs.}
\label{fig:DeepESNtrain}
\end{figure}

A deep ESN is an ESN composed of $L$ stacked reservoirs, as shown in the sketch of the deep ESN training phase in Fig.~\ref{fig:DeepESNtrain}. 

In this case, $W^{(l)}$ denotes the $l$th reservoir weight matrix, $W^{\mathrm{in}(l)}$ the $l$th input weight matrix, $x_k^{(l)}$ the local internal reservoir state vector, and $x_k$ the global internal reservoir state vector. Equations \eqref{eq:discrESN} and \eqref{eq:discrESN2} for a shallow ESN read now
\begin{align}
    x_{k}^{(l)} & = \left(1-a \Delta t \right)x_{k-1}^{(l)} + \Delta t f(W^{(l-1)}x_{k}^{(l-1)} + W^{(l)}x_{k-1}) \qquad l>1 \nonumber
    \\
    & \label{eq:discrDeepESN} \\
    x_{k}^{\mathrm{out}} & = g(W^{\mathrm{out}}[x_{k};u_k]) \, , \nonumber
\end{align}
where $x_k$ is the concatenation of all $x_k^{(l)}$.

\section{ESN predictive model for DA evolution}\label{sec4}

In the previous section, we have introduced the definition of a shallow leaky ESN and its extension as a deep ESN. In Eqs.~\eqref{eq:discrESN} and~\eqref{eq:discrDeepESN} we can already identify some parameters (called hyperparameters) of the ESN predictive model. 
These are the leaking rate $a$, the number of stacked reservoirs $L$, the dimension $N_\mathrm{r}$ of the reservoir matrix $W$ 
 and the activation  function $f$ usually set as the hyperbolic tangent function $\tanh$. In Appendix~\ref{AP:ESP}, we give a sufficient condition on the spectral radius $\rho$ of the reservoir matrix $W$,  which can also be considered as a hyperparameter, that guarantees the Echo State Property (ESP). 

Other hyperparameters are often introduced in the implementation of ESN equations. Specifically, the sparsity ratio $s$ of the reservoir matrix $W$, i.e. the fraction of 0 elements in the reservoir matrix $W$ and $BI$ (as in~\cite{topology}), which corresponds to the number of time steps of the input data that are discarded. Furthermore, the regularisation parameter $\beta$ in Eq.~\eqref{eq:ridge} also needs to be optimised and is also considered a hyperparameter of the ESN model. Setting large values for $\beta$ is generally used to avoid overfitting and may improve prediction in the \textit{test set}. To complete the definition of the predictive model of the ESN, we must assign a value to all hyperparameters, knowing that the performance of the model strongly depends on the choice of their values.

It is a common procedure in ESN training to perform an optimisation of these hyperparameters, which is usually done by grid search methods~\cite{algosearch}, in the \textit{validation set}. The validation procedure considered here is based on an ensemble approach to deal with the randomness of the reservoirs. Eventually, once the predictive model has been trained and validated, we can test it in the \textit{test set} with unseen data.

\subsection{ESN ensemble validation approach}

The ensemble validation approach used in our studies is based on the principle of minimising the average of the Relative Root Mean Square Error (RRMSE) of $N_\mathrm{d}$ dynamics  predicted (i.e, 60 seeds for the HL-LHC dataset and 60 cases for the 4D Henon map) for $N_\mathrm{W}$ different randomly generated reservoirs and various hyperparameters values on the \textit{validation set}. Note that for each of the $N_\mathrm{d}$ dynamics, we predict a mean over the $N_\mathrm{W}$ reservoirs. Additionally, each of the $N_\mathrm{d}$ dynamics contains different input/training/validation/test data, so that each prediction is performed independently of the others. We define this RRMSE on the \textit{validation set}  $\mathrm{RRMSE^{val}}$ as:
\begin{align}
   \mathrm{RRMSE^{val}} = \frac{1}{N_\mathrm{d}}\sum_{i=1}^{N_\mathrm{d}} \left ( 100 \sqrt{\frac{\sum_{k=1}^{k_{\mathrm{val}}} (x_{\mathrm{mean},k}^{\mathrm{out}-i} - x_k^{\mathrm{val}-i})^2}{\sum_{k=1}^{k_{\mathrm{val}}} (x_k^{\mathrm{val}-i})^2}} \right )
\end{align}
where $k_{\mathrm{val}}$ is the number of validation data, $x_{\mathrm{mean}}^{\mathrm{out}-i}$ is the mean over the $N_\mathrm{W}$ reservoirs for the $i$th dynamics at time $k$, and $x_{k}^{\mathrm{val}-i}$ is the validation data at the same time $k$ for the same $i$th dynamics.

This procedure aims to build a robust predictive model in which all hyperparameters are fixed. The search of the hyperparameters values minimizing the $\mathrm{RRMSE^{val}}$ is done over a domain $S_h$. Each of the hyperparameters is updated one by one using the value in $S_h$, which minimises $\mathrm{RRMSE^{val}}$. Furthermore, as mentioned above, this ensemble validation method requires the generation of different random matrices $W$ and $W^{\mathrm{in}}$. This is done by sampling their elements from a uniform pseudorandom distribution in $(0,1)$ and scaling them to the interval (-0.5,0.5) so that they also have negative elements. 
The procedure for generating $W^{\mathrm{in}}$ and $W$ is detailed in Algorithm \ref{algo:1}, while a pseudocode of the general ensemble validation procedure is presented in Algorithm \ref{algo:2}. 
\begin{algorithm}
\caption{Generation of the random matrix $W_{\text{in}}$ and $W$.}\label{algo:1}
\begin{algorithmic}[1]
\Require $\rho$  spectral radius we want to set to $\Tilde{W} = \Delta t \| W \|+(1 \text{-}a \Delta t)I$, $K$  input size, $N_\mathrm{r}$  reservoir size
\Ensure $W^{\mathrm{in}}$ input weight matrix, $W$ reservoir weight matrix
\State Random initialisation of $W^{\mathrm{in}}_{i,j}$ $\sim$ $\mathcal{U}$ (0,1) - 0.5 , $i = 1, \ldots, N_\mathrm{r}$, $j = 1, \ldots, K$, $W_{i,j}$ $\sim$ $\mathcal{U}$ (0,1) - 0.5 , $i,j = 1, \ldots, N_\mathrm{r}$
\State  Compute spectral radius $\rho_{\mathrm{rand}}$ of $\Tilde{W}$ 
\State Scale W = $\rho/\rho_{\mathrm{rand}}$ W
\end{algorithmic}
\end{algorithm}

\begin{algorithm}
\caption{Validation.}\label{algo:2}
\begin{algorithmic}[1]
\Require $S_h$ domain of search of an hyperparameter $h$, $N_\mathrm{W}$ random different pairs of  ($W^\mathrm{in}$,$W$), $N_\mathrm{d}$ dynamics we want to validate with the associated inputs data $u$, training data $x^{\mathrm{train}}$ and validation data $x^{\mathrm{val}}$.
\Ensure  $H$ set of all the fixed hyperparameters $h$ ($N_\mathrm{r}$, $L$, $BI$ $\rho$, $\beta$,$\Delta t$)  which minimise in average RRMSE$^{\mathrm{val}}$ for the $N_\mathrm{d}$ dynamics and $N_\mathrm{W}$ reservoirs.

\For {$h \in H $}
\For{$h_{val} \in S_{h}$}
\For {$i$ = $1$ to $N_{d}$} 
\State $x_{mean}^{\mathrm{out}-i}$ = 0
    \For {$j$ = $1$ to $N_\mathrm{W}$} 
        \State $W^{\mathrm{out}}$ = Training($u^{-i}$, $x^{\mathrm{train}-i}$, $h_{val}$, $(W^{\mathrm{in}},W)^j$)
        \State $x^{\mathrm{out}-i}$ = Prediction($h_{val}$, $(W^{\mathrm{in}},W)^j$, $W^{\mathrm{out}}$)
        \State $x_{\mathrm{mean}}^{\mathrm{out}-i}$+=$x^{\mathrm{out}-i}$
    \EndFor
        \State  $x_{\mathrm{mean}}^{\mathrm{out}-i}$ = $x_{mean}^{\mathrm{out}-i}$ / $N_\mathrm{W}$
        \State $\mathrm{RRMSE}^{\mathrm{val}}_{h_{val}}$ += $100 \sqrt{\frac{\sum_{k=1}^{k_{\mathrm{val}}} (x_{\mathrm{mean},k}^{\mathrm{out}-i} - x_k^{\mathrm{val}-i})^2}{\sum_{k=1}^{k_{\mathrm{val}}} (x_k^{\mathrm{val}-i})^2}}$
\EndFor
\State $\mathrm{RRMSE}^{\mathrm{val}}_{h_{val}}$ =  $\mathrm{RRMSE}^{\mathrm{val}}_{h_{val}} / N_d$
\EndFor
\State $h_{val}$ = arg min($\mathrm{RRMSE}^{\mathrm{val}}_{h_{val}}$)
\State Set $h$ := $h_{val}$ and update $H$
\EndFor
\end{algorithmic}
\end{algorithm}

Note that the functions Training() and Prediction() implement the equations presented in Section~\ref{sec3}. 

\subsection{ESN ensemble test approach}

Once the parameters and hyperparameters of the ESN predictive model have been tuned using \textit{training set} and \textit{validation set}, we can test our ESN model for the prediction of not previously used data, i.e. DA values at a larger time. We denote by $k_{\mathrm{test}}$ the number of data in the \textit{test set} we try to predict.
\begin{algorithm}
\caption{Test} \label{algo:3}
\begin{algorithmic}[1]
\Require  $N_\mathrm{W}$ number of different pairs of  ($W^{\mathrm{in}}$,$W$), $x^{\mathrm{test}}$ test  data of size $k_{\mathrm{test}}$, \textit{H} = $\{N_\mathrm{r}$, $\Delta t$, $\rho$, $a$,  $BI$, $L$, $\beta \}$ have been tuned using the $k_{\mathrm{val}}$ data in the \textit{validation set} with the corresponding output weight matrix $W^{\mathrm{out}}$.
\Ensure $x_{\mathrm{mean}}^{\mathrm{out}}$ mean of the predicted outputs over the $N_\mathrm{W}$ reservoirs, $\mathrm{RRMSE}^{\mathrm{test}}$ between  $x_{\mathrm{mean}}^{\mathrm{out}}$ and $x^{\mathrm{test}}$
\State $x_{\mathrm{mean}}^{\mathrm{out}}$ = 0
\For{$j$ = $1$ $to$ $N_\mathrm{W}$}
    \State $x^{\mathrm{out}}$ = Prediction($H$, $(W^{\mathrm{in}},W)^j$, $W^{\mathrm{out}}$)
    \State $x_{\mathrm{mean}}^{\mathrm{out}}$ += $x^{\mathrm{out}}$
\EndFor
\State  $x_{\mathrm{mean}}^{\mathrm{out}}$ /= $N_\mathrm{W}$
\State $\mathrm{RRMSE}^{\mathrm{test}}$ = $100 \sqrt{\frac{\sum_{k=1}^{k_{\mathrm{test}}} (x_{\mathrm{mean},k}^{\mathrm{out}} - x_k^{\mathrm{test}})^2}{\sum_{k=1}^{k_{\mathrm{test}}} (x_k^{\mathrm{test}})^2}}$
\end{algorithmic}
\end{algorithm}

The algorithm~\ref{algo:3} describes the test procedure for a single dynamics, i.e. a single realisation of the HL-LHC magnetic lattice or a single case for the 4D H\'enon map data set. We can loop the procedure to perform the prediction in the \textit{test set} for the $N_\mathrm{d}$ dynamics. Note that, contrary to the validation, here the prediction is performed in the \textit{test set} for data not previously used.

\section{Results and Discussion}\label{sec5}

In this section, we present the DA predictions obtained with our ESN-based predictive model. In particular, we compare these predictions with those of the fitted scaling law presented in Eq.~\eqref{model2.1_1} and used in~\cite{ScalingLaws}. We recall that the ESN output $x_{\mathrm{mean}}^{\mathrm{out}}$ is the mean prediction over $N_\mathrm{W} = 100$ random reservoirs. The validation and testing methods are those introduced in Section~\ref{sec4}. We tested the proposed approaches with the HL-LHC data sets and the 4D H\'enon map presented in Section~\ref{sec2}.

\subsection{DA Predictions for the HL-LHC data set}

\subsubsection{Validation of the ESN}

In this stage, we search for the set of hyperparameters $H$ that minimises, on average over the $N_\mathrm{d}=60$ seeds and $N_\mathrm{W}=100$ randomly generated reservoirs, the RRMSE in the \textit{validation set}. Here, the number of predicted dynamics is equal to the number of seeds. We also recall that the number of validation data is $k_{\mathrm{val}}= 50$ and the definition of $\mathrm{RRMSE}^{\mathrm{val}}$ is presented in Algorithm \ref{algo:2}. The optimal hyperparameters are determined one by one by a grid search over a wide range of possible parameter values, and the search domains $S_h$ of the hyperparameters are listed in Table~\ref{tab:hldom}. 

\begin{table}[htb]
\centering
\caption{Search domains $S_h$ of the various hyperparameters $h$. Note that the hyperparameters $h$ are tuned one by one while the others are kept constant.}
\begin{tabular}{|c|c|} 
\hline
 $h$ & $S_h$  \\ 
\hline
$N_\mathrm{r}$& $\left \{20, 40, 60, \ldots , 200 \right \}$ \\
\hline
$L$ & $\left \{1, 2, 3 \right \}$ \\ 
\hline
$\rho$ & $\left \{0.1, 0.2, \ldots , 0.9, 0.99 \right \}$ \\ 
\hline
$BI$ & $\left \{0,25,50, \ldots , 200 \right \}$ \\ 
\hline
$\beta$ & $\left \{2 \times 10^{-10}, 2 \times 10^{-8}, \ldots , 2 \times 10^{-2}, 2 \times 10^{-1} \right \}$ \\ 
\hline
$\Delta t$ & $\left \{9 \times 10^{-7}, 9 \times 10^{-6}, \ldots , 9 
\times 10^{-2}, 9 \times 10^{-1}\right \}$ \\ 
\hline
\end{tabular}
\label{tab:hldom}
\end{table}

\begin{figure}[!htb]
\minipage{0.5\textwidth}
\centering
  \includegraphics[width=5.3cm]{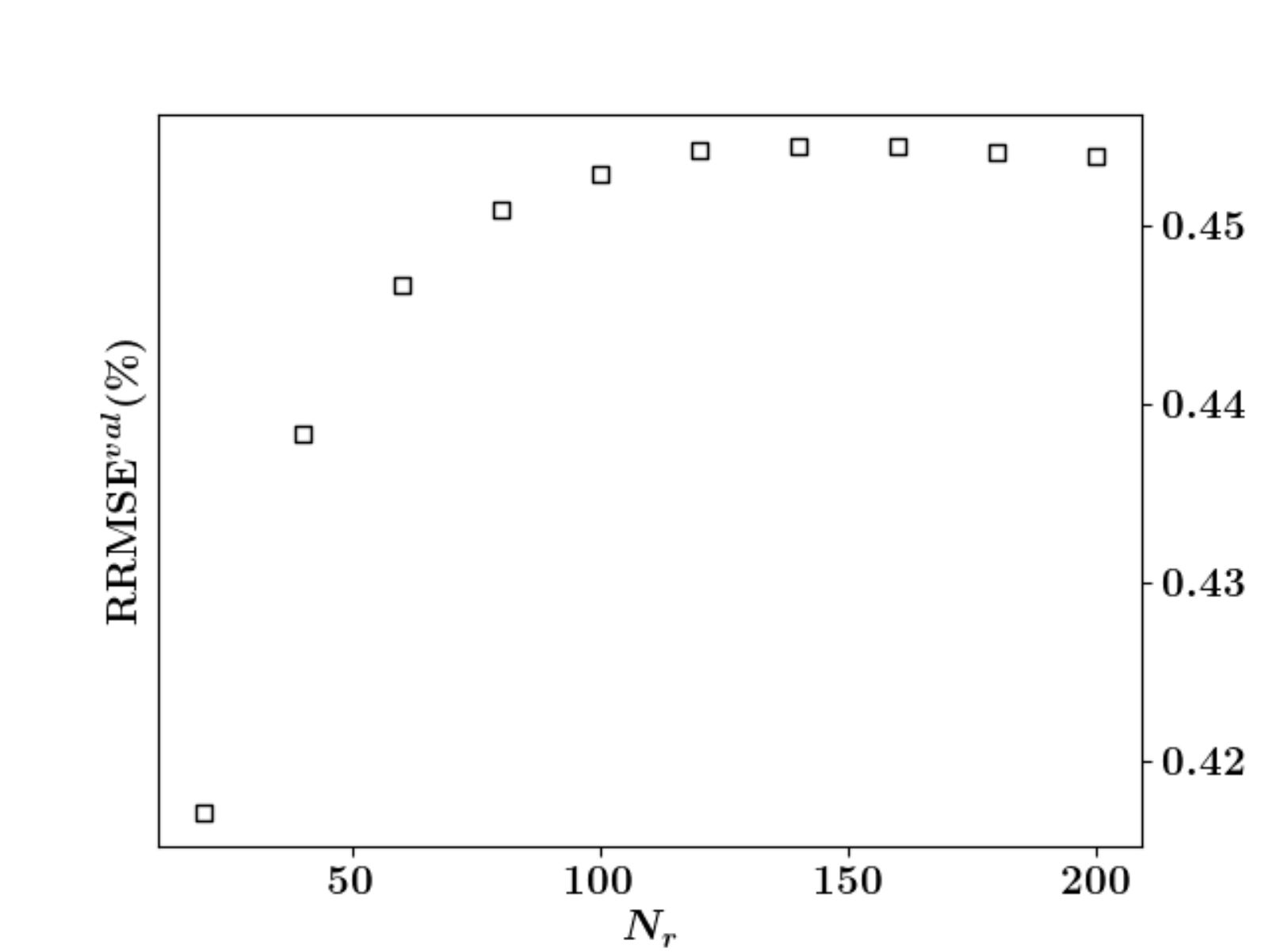}
  \caption*{\small{Values of constant hyperparameters: $L=1$, $\rho=0.1$, $BI=2\times 10^{2}$, $\beta=2\times 10^{-1}$, $\Delta t=9 \times 10^{-2}$.}} 
\endminipage\hfill
\minipage{0.5\textwidth}
\centering
  \includegraphics[width=5.3cm]{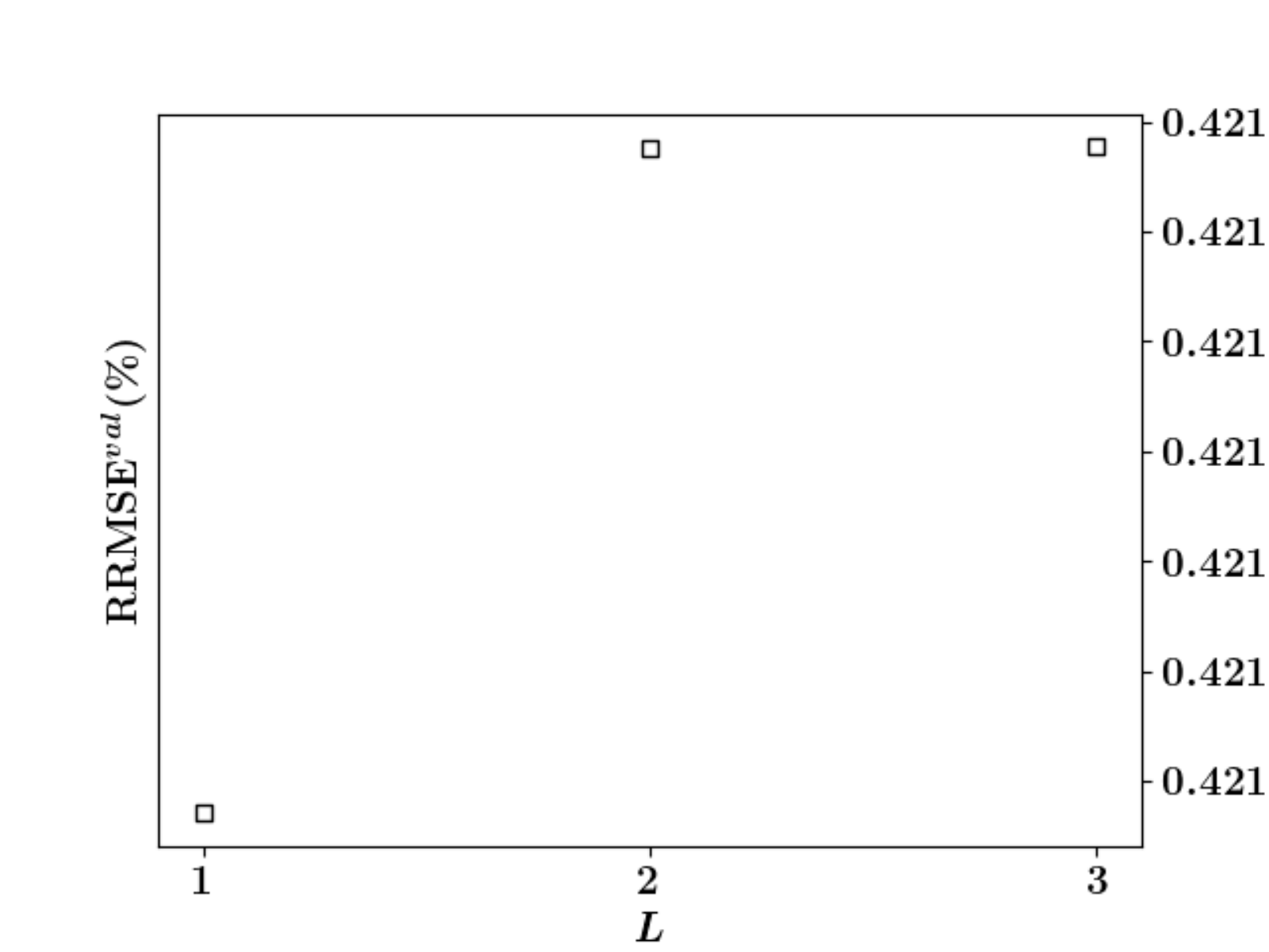}
   \caption*{\small{Values of constant hyperparameters:  $N_\mathrm{r}=20$ $\rho=0.1$,  $BI=2\times 10^{2}$, $\beta=2\times 10^{-1}$, $\Delta t=9 \times 10^{-2}$.}}
\endminipage\hfill
\\
\minipage{0.5\textwidth}
\centering
  \includegraphics[width=5.3cm]{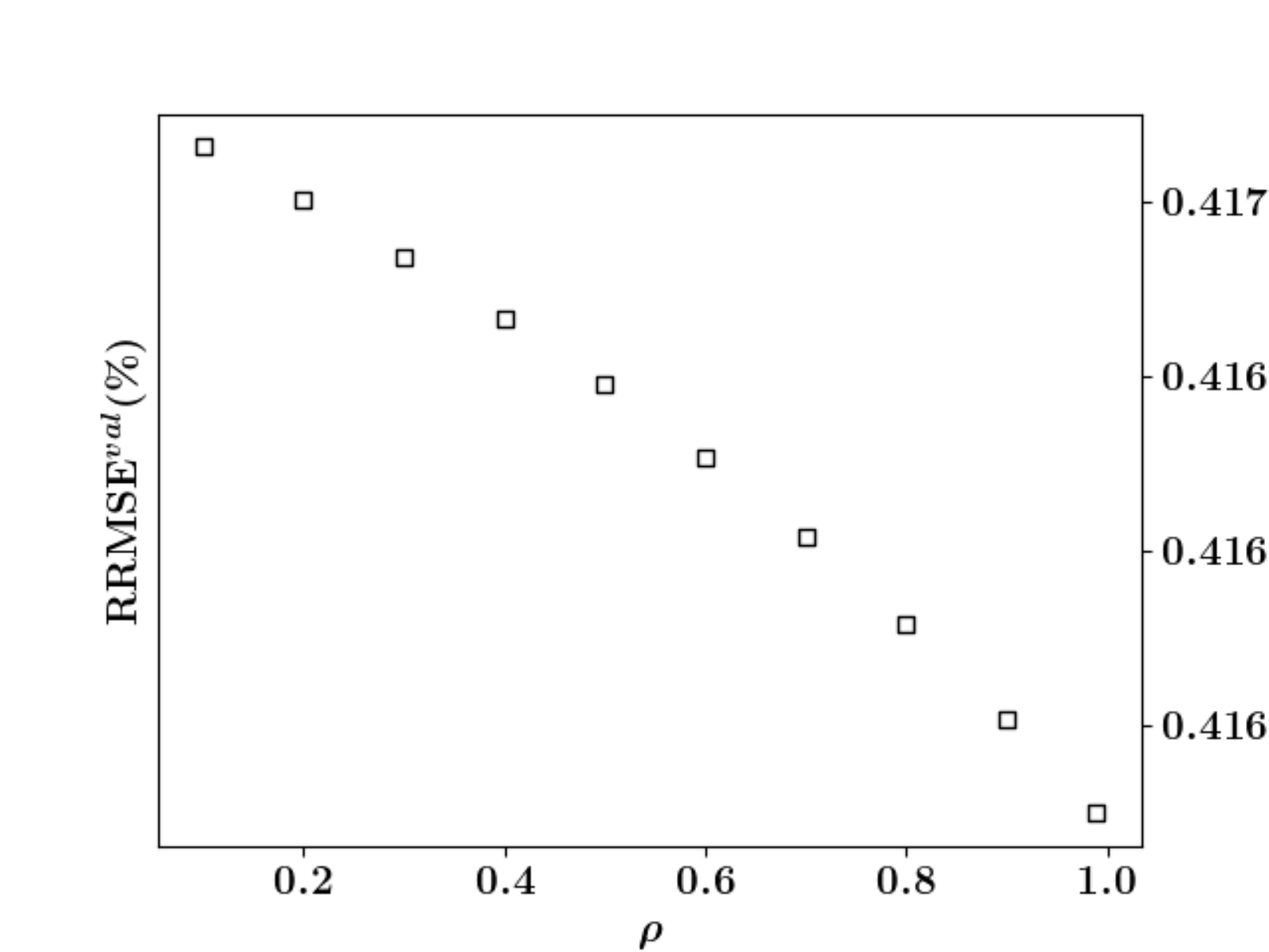}
  \caption*{\small{Values of constant hyperparameters: $N_\mathrm{r}=20$, $L=1$, $BI=2\times 10^{2}$, $\beta=2\times 10^{-1}$, $\Delta t=9 \times 10^{-2}$.}}
\endminipage\hfill
\minipage{0.5\textwidth}
\centering
  \includegraphics[width=5.3cm]{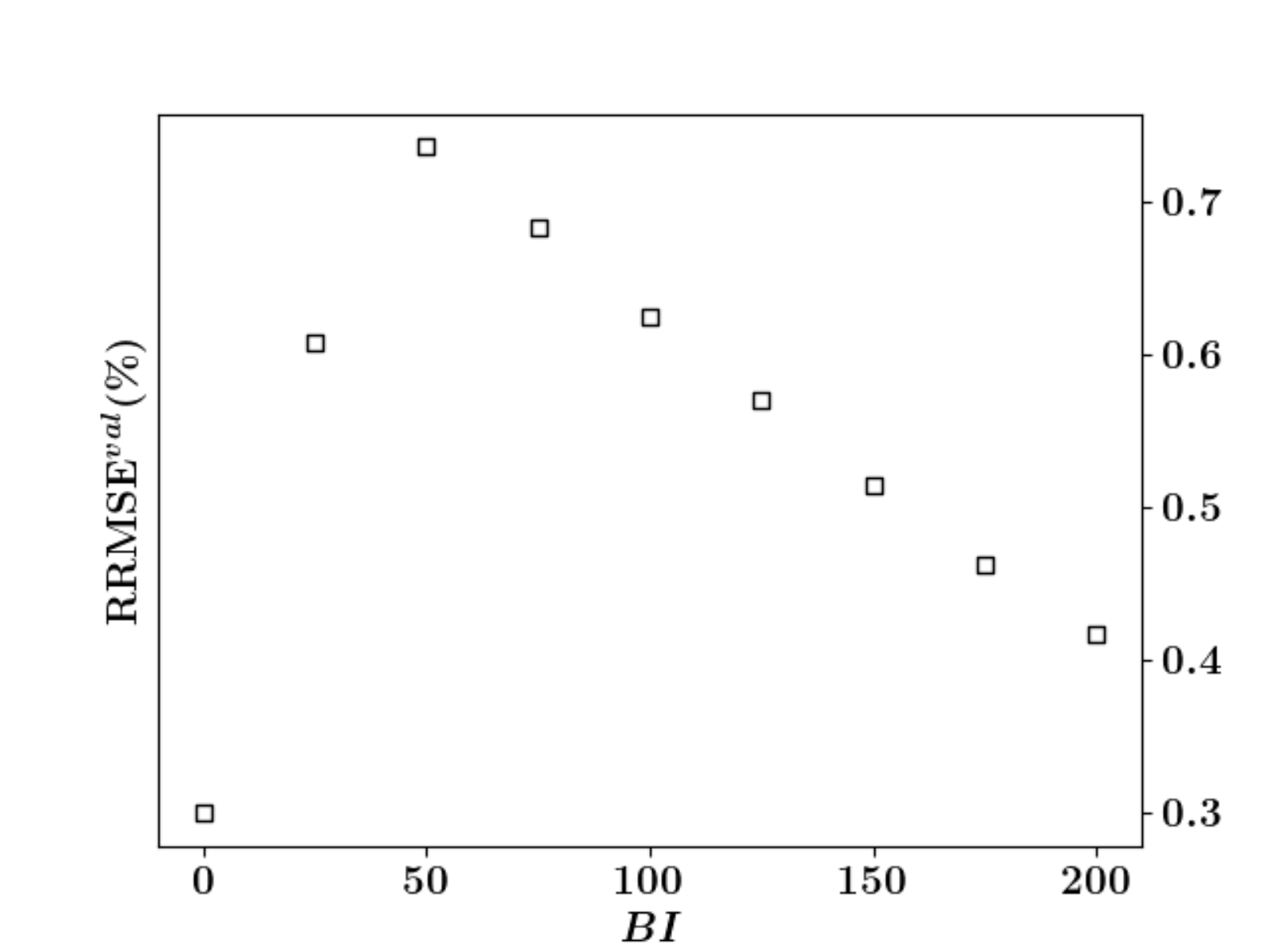}
  \caption*{\small{Values of constant hyperparameters: $N_\mathrm{r}=20$, $L=1$, $\rho=0.99$, $\beta=2\times 10^{-1}$, $\Delta t=9 \times 10^{-2}$.}}
\endminipage\hfill
\minipage{0.5\textwidth}
\centering
  \includegraphics[width=5.3cm]{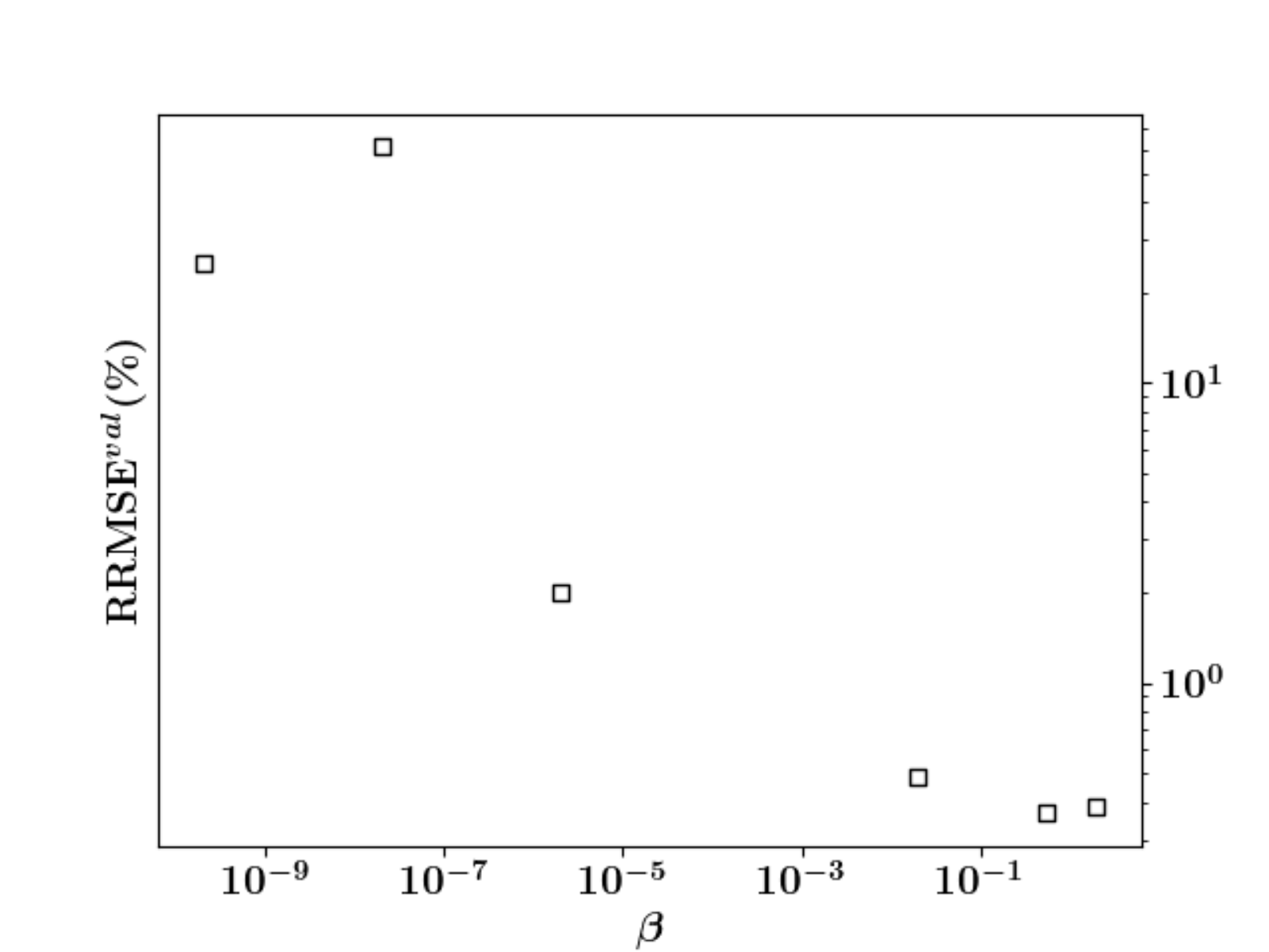}
  \caption*{\small{Values of constant hyperparameters: $N_\mathrm{r}=20$, $L=1$, $\rho=0.99$, $BI=0$, $\Delta t=9 \times 10^{-2}$.}}
\endminipage\hfill
\minipage{0.5\textwidth}
\centering
  \includegraphics[width=5.3cm]{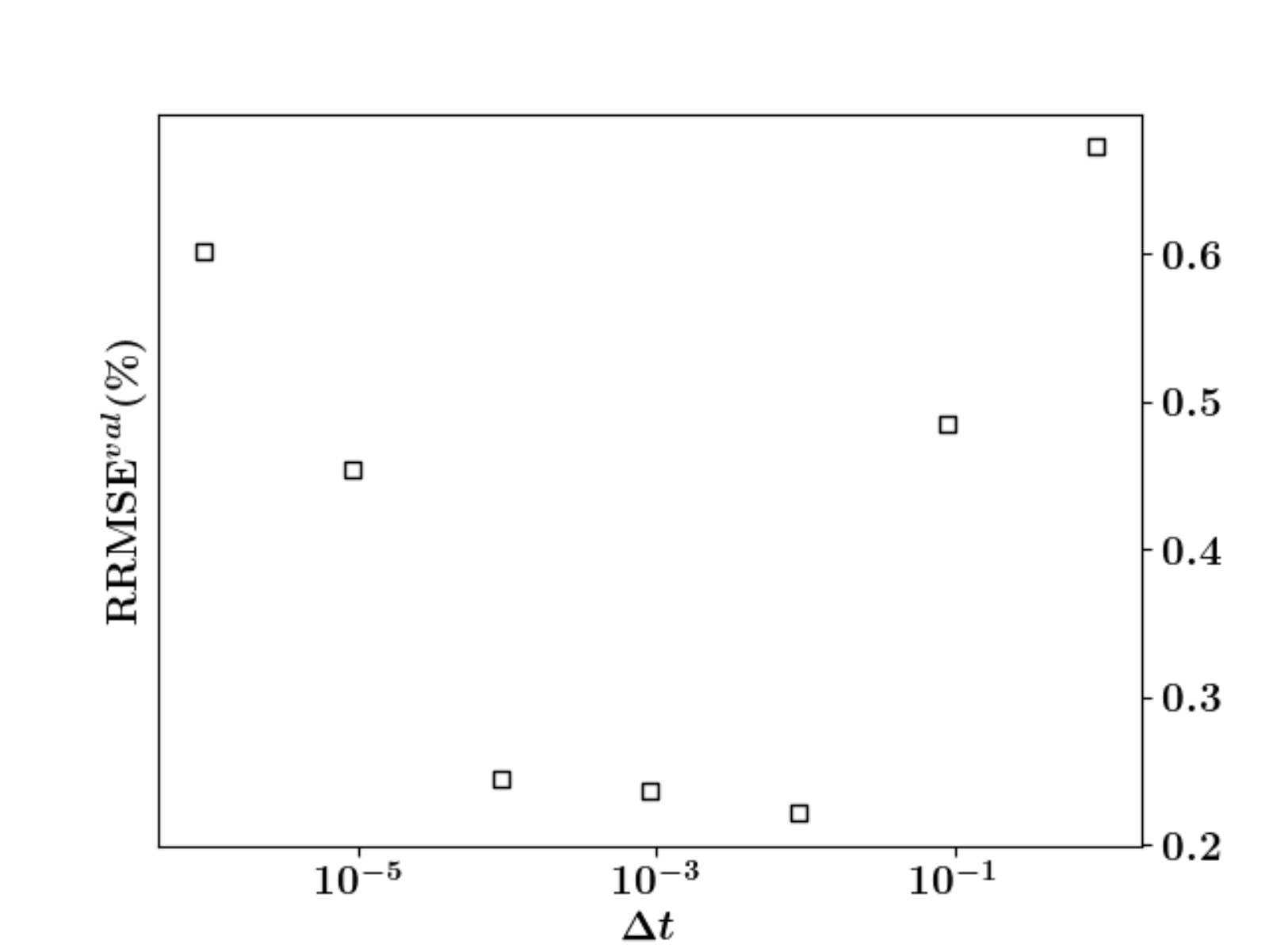}
  \caption*{\small{Values of constant hyperparameters: $N_\mathrm{r}=20$, $L=1$, $\rho=0.99$, $BI=0$, $\beta=2 \times 10^{-2}$.}}
\endminipage\hfill
\caption{$\mathrm{RRMSE^{val}}$ as a function of the various hyperparameters in $S_h$.}
\label{fig:hlsens}
\end{figure}

Figure~\ref{fig:hlsens} shows $\mathrm{RRMSE^{val}}$ as a function of the various hyperparameters in $S_h$.  The values of the hyperparamters are updated one-by-one with those that minimise RMMSE$^{val}$.  
As we can see, a shallow ESN with a small number of neurons $N_\mathrm{r}$ provides the best results. Stacking more reservoirs does not improve the predictions. In fact, adding reservoirs or increasing the number of neurons makes the model overfit, so it cannot predict correctly in the \textit{validation set}. This can be explained by the small number of features that the ESN must learn and by the characteristics of the DA data, which are not enough. 

Regarding the other hyperparameters, the optimum spectral radius value initially set to 0.1 is updated to 0.99 and satisfies the ESP. Furthermore, since the optimal value of $N_\mathrm{r}$ is smaller than $100$, it can be considered small, which justifies setting the sparsity ratio $s$ = 0 so that all elements of $W$ are non-zero. Then, we decided to choose the activation function $f = \tanh$, since it is the most used in ESN, and the leaking rate $a = 1$ to simplify the equations described in~\eqref{eq:discrDeepESN}. Eventually, the values of $\beta$ and $\Delta t$ initially set to $2.10^{-1}$ and $9.10^{-2}$ are updated to $2.10^{-2}$ and $9.10^{-3}$ respectively. The values of the hyperparameters updated after validation and used for the prediction stage in the \textit{test set} are summarised in Table~\ref{tab:hlsens}.

\begin{table}[htb]
\centering
\caption{Set $H$ of the hyperparameters tuned after validation using HL-LHC DA data.}
\begin{tabular}{|c|c|c|c|c|c|c|c|c|} 
\hline
$N_\mathrm{r}$ & $\beta$ & $\rho$ & $a$ & $BI$ & $L$ & $\Delta t$  & $f$ & $s$\\
\hline
\hline
20 & $2.10^{-2}$ & 0.99 & 1 & 0 & 1 & $9.10^{-3}$ & $\tanh$& 0 \begin{tabular}[c]{@{}c@{}}\end{tabular} \\
\hline
\end{tabular}
\label{tab:hlsens}
\end{table}

\subsubsection{The ESN model}

\begin{figure}[!htb]
\minipage{0.5\textwidth}
\centering
  \includegraphics[width=5.1cm]{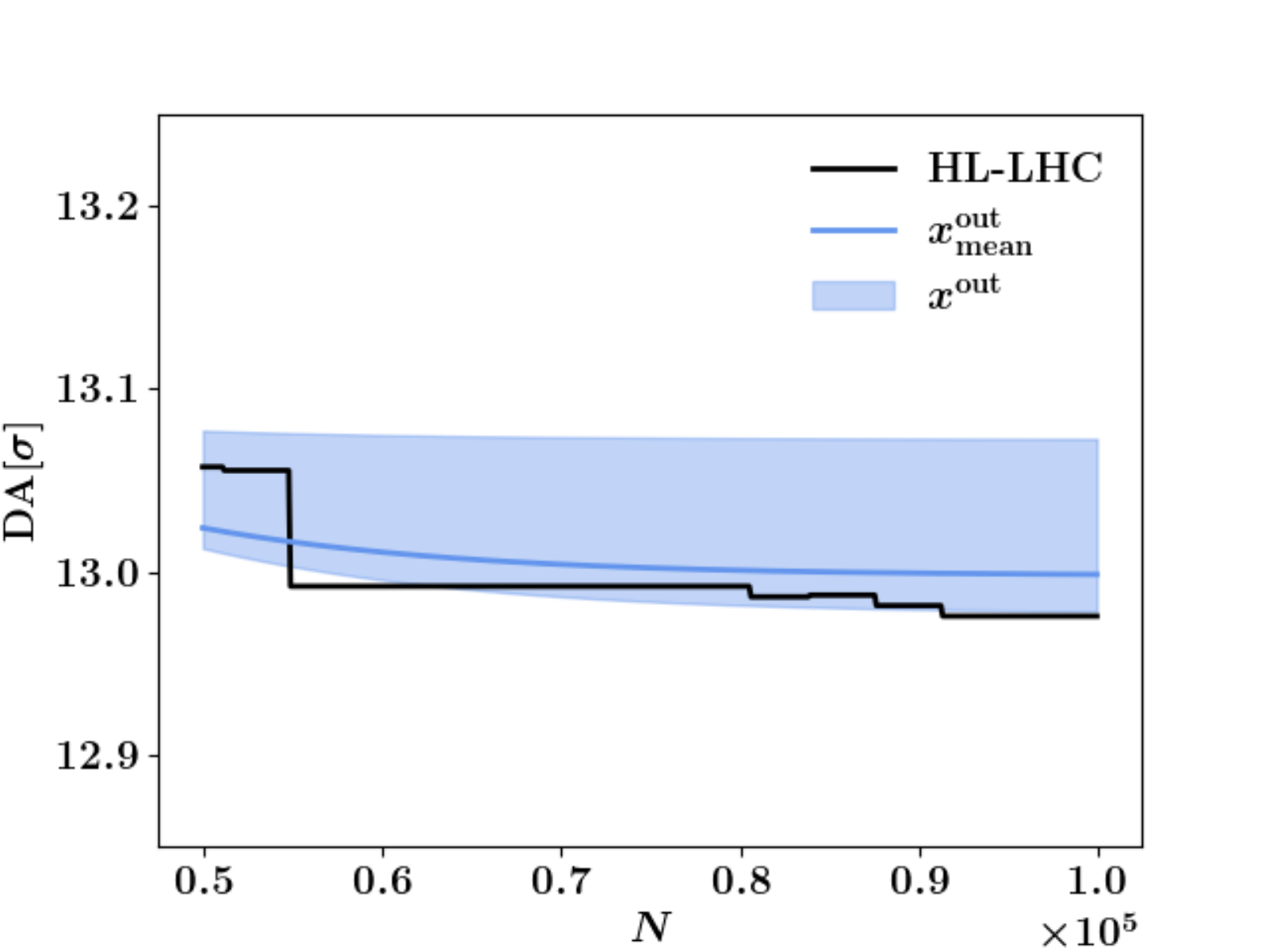}
\endminipage\hfill
\minipage{0.5\textwidth}
\centering
  \includegraphics[width=5.3cm]{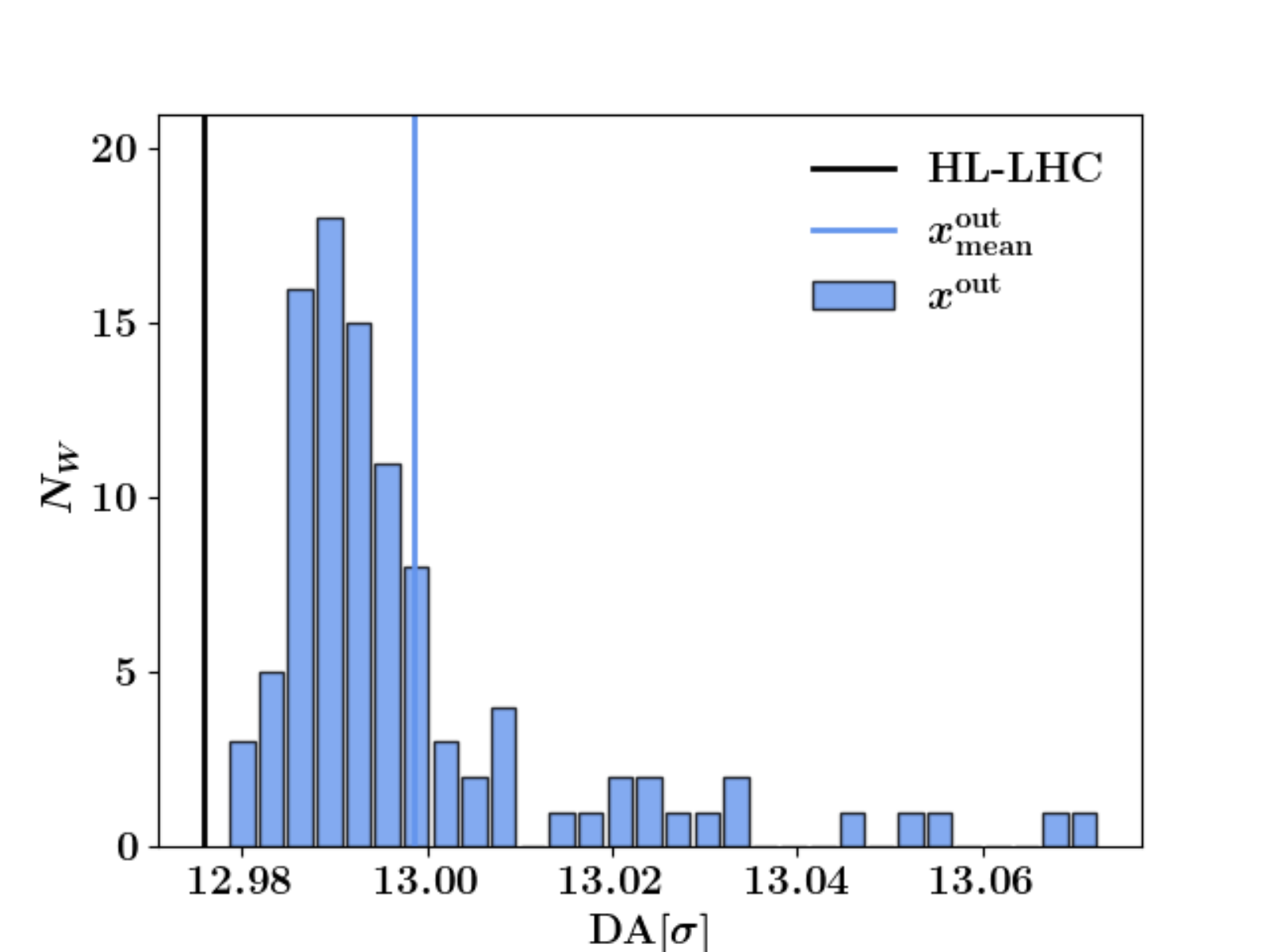}
\endminipage\hfill
\caption{Left: Numerical DA data, prediction of DA $x_{\mathrm{mean}}^{\mathrm{out}}$, average, minimum, and maximum over the $N_\mathrm{W} = 100$ randomly generated reservoirs as a function of time. Right: distribution for the $N_\mathrm{W} = 100$ randomly generated reservoirs at $N=10^5$. The seed used for both plots is number 1.}
\label{fig:hlarb}
\end{figure}

Once the ESN has been trained and validated, we can test it with the $\textit{test set}$ for data not previously used using the hyperparameters reported in Table~\ref{tab:hlsens}. We recall that the number of test data is $k_{\mathrm{test}} = 500$, i.e. half of the total number of data used. In Fig.~\ref{fig:hlarb}, we show the mean prediction $x_{\mathrm{mean}}^{\mathrm{out}}$ in the \textit{test set} together with the envelope (i.e. minimum and maximum) of the predictions $x^{\mathrm{out}}$ that are associated with the $N_\mathrm{W} = 100$ randomly generated reservoirs for an arbitrary seed (number 1). We also plot the distribution of the prediction of DA at $N = 10^5$ turns (end of the $\textit{test set}$).

\begin{figure}[!htb]
\minipage{0.5\textwidth}
\centering
  \includegraphics[width=5.1cm]{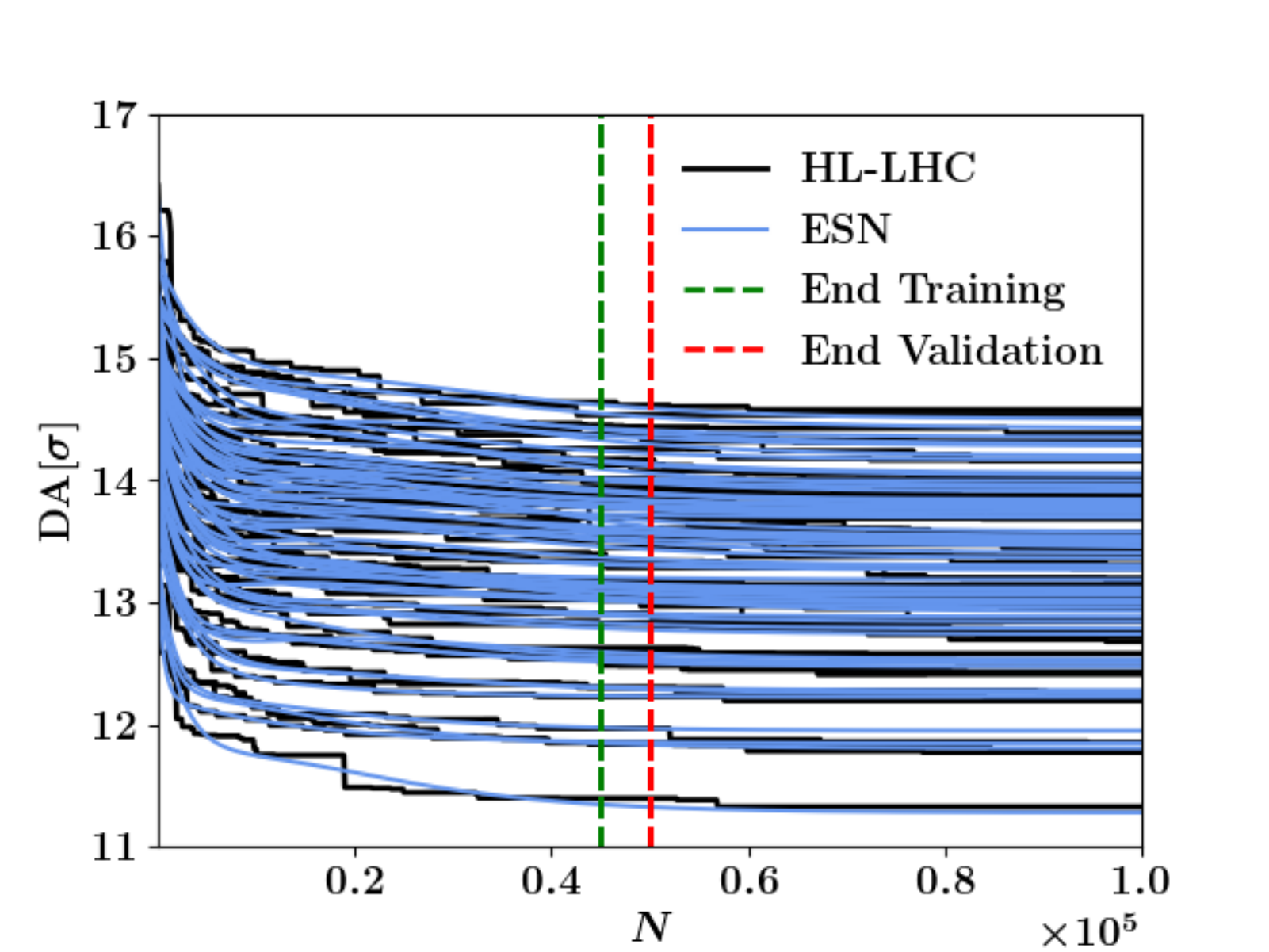}
\endminipage\hfill
\minipage{0.5\textwidth}
\centering
  \includegraphics[width=5.3cm]{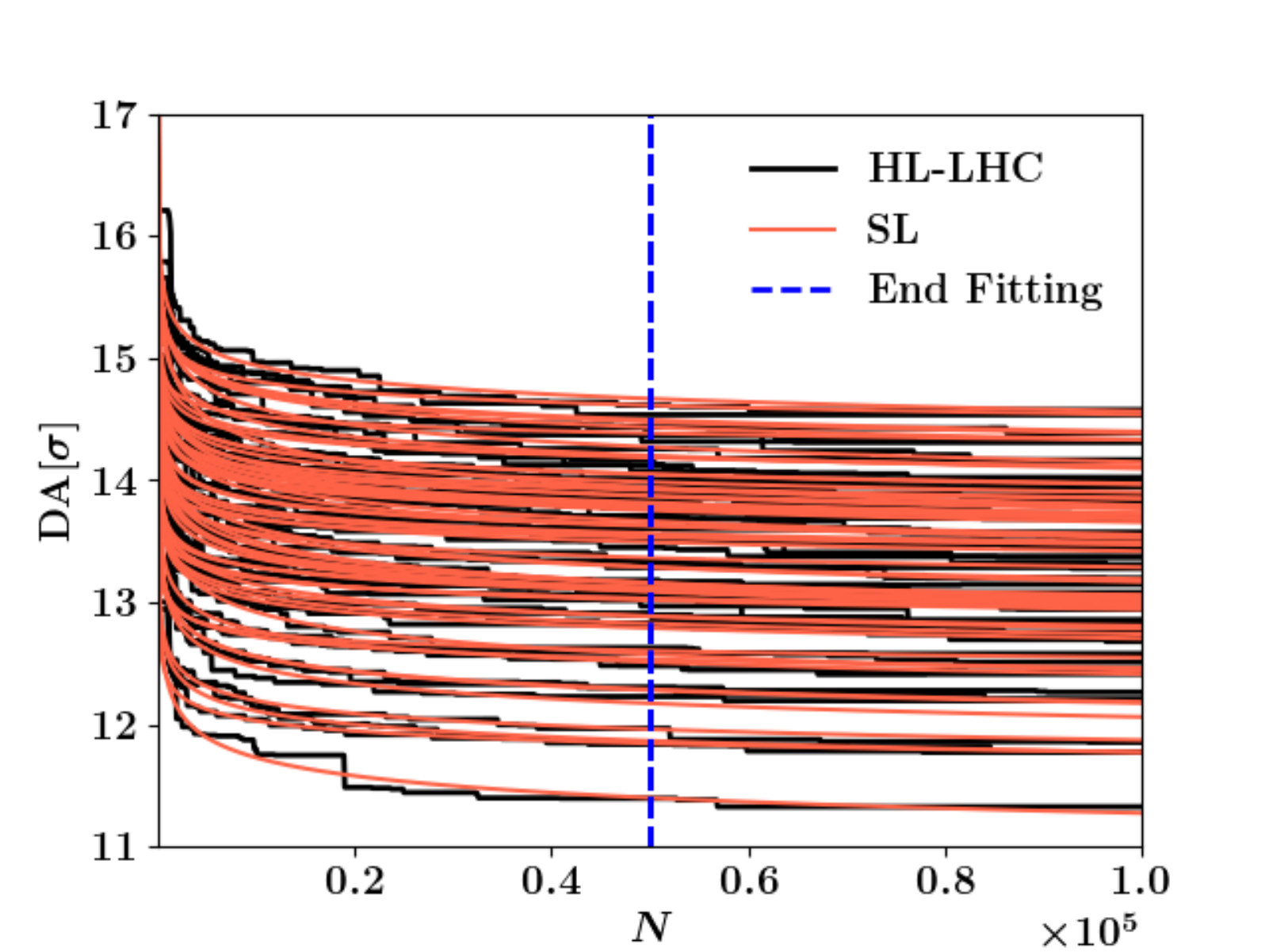}
\endminipage\hfill
\caption{DA predictions for ESN (left) and SL (right) for $N_\mathrm{d}$ = 60 seeds.}
\label{fig:hlpred}
\end{figure}

\begin{figure}[!htb]
\centering
  \includegraphics[width=5.3cm]{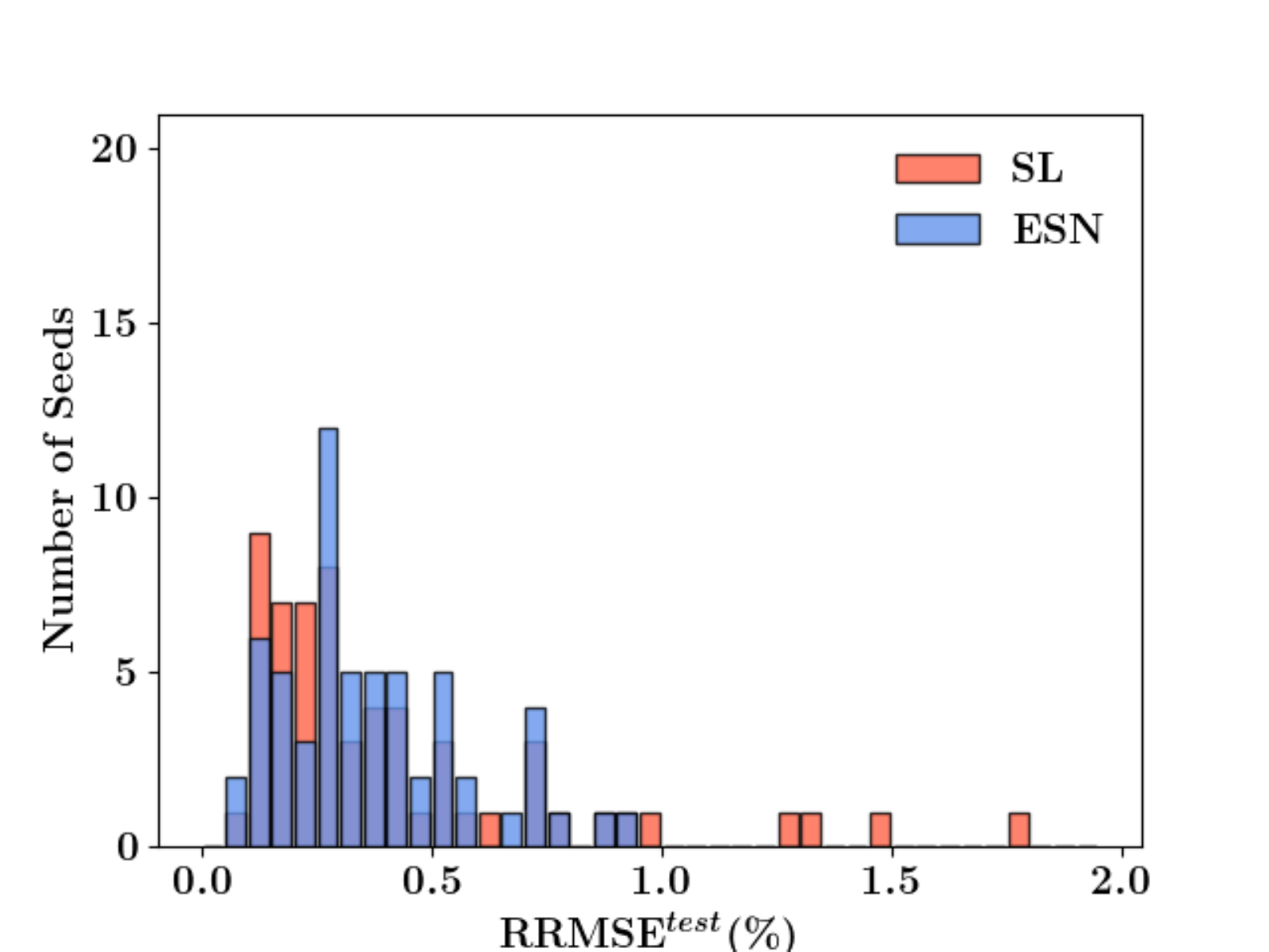}
\caption{Distribution of $\mathrm{RRMSE^{test}}$ for $N_\mathrm{d}$=60 seeds for ESN and SL.}
\label{fig:hlrrmse}
\end{figure}

As mentioned above, we will denote by  $x_{\mathrm{mean}}^{\mathrm{out}}$ the ESN mean prediction and only plot this mean value to avoid overloading the graphs with the values generated by $N_\mathrm{W}$ random reservoirs. To have a complete view, Fig.~\ref{fig:hlpred} shows the predictions of $N_\mathrm{d} = 60$ seeds in the \textit{train set}, \textit{validation set} and \textit{test set}. Vertical dashed lines indicate the end of the \textit{train set} and \textit{validation set} for ESN (left graph) and SL (right graph). The scaling law fit is performed using the first $k_{\mathrm{fit}} = k_{\mathrm{train}}+k_{\mathrm{val}}=500$ DA data. Note that ESN and SL share the same \textit{test set}. Figure~\ref{fig:hlrrmse} shows the distribution of the $\mathrm{RRMSE^{test}}$ values defined in Algorithm \ref{algo:3}, for both the ESN model and SL.

We report in Table~\ref{tab:hlcompa} the mean, maximum, minimum, and standard deviation of $\mathrm{RRMSE}^{test}$ for the predictions of ESN and SL over $N_\mathrm{d}$ = 60 seeds.

\begin{table}[htb]
\centering
\caption{Mean, maximum, minimum, and standard deviation of the $\mathrm{RRMSE^{test}}$ distribution.}
\begin{tabular}{|c|c|c|c|c|c|c|c|c|} 
\hline
 &  Mean & Max & Min & Std \\
\hline
\hline
ESN & 0.37 & 0.94 & 0.06 & 0.20 \begin{tabular}[c]{@{}c@{}}\end{tabular} \\
\hline
SL & 0.42 & 1.78 & 0.07 &  0.35 \begin{tabular}[c]{@{}c@{}}\end{tabular} \\
\hline
\end{tabular}
\label{tab:hlcompa}
\end{table}

The ESN model and SL generate predictions whose distributions have essentially the same mean and minimum values. However, some outliers appear in the SL distribution, which affect the maximum and standard deviation values. This contributes to the generation of more stable predictions by ESN, i.e. without outliers, and significantly lower values of the standard deviation and maxima.

\subsubsection{The SL-ESN model}

\begin{figure}[!htb]
\minipage{0.5\textwidth}
\centering
  \includegraphics[width=5.3cm]{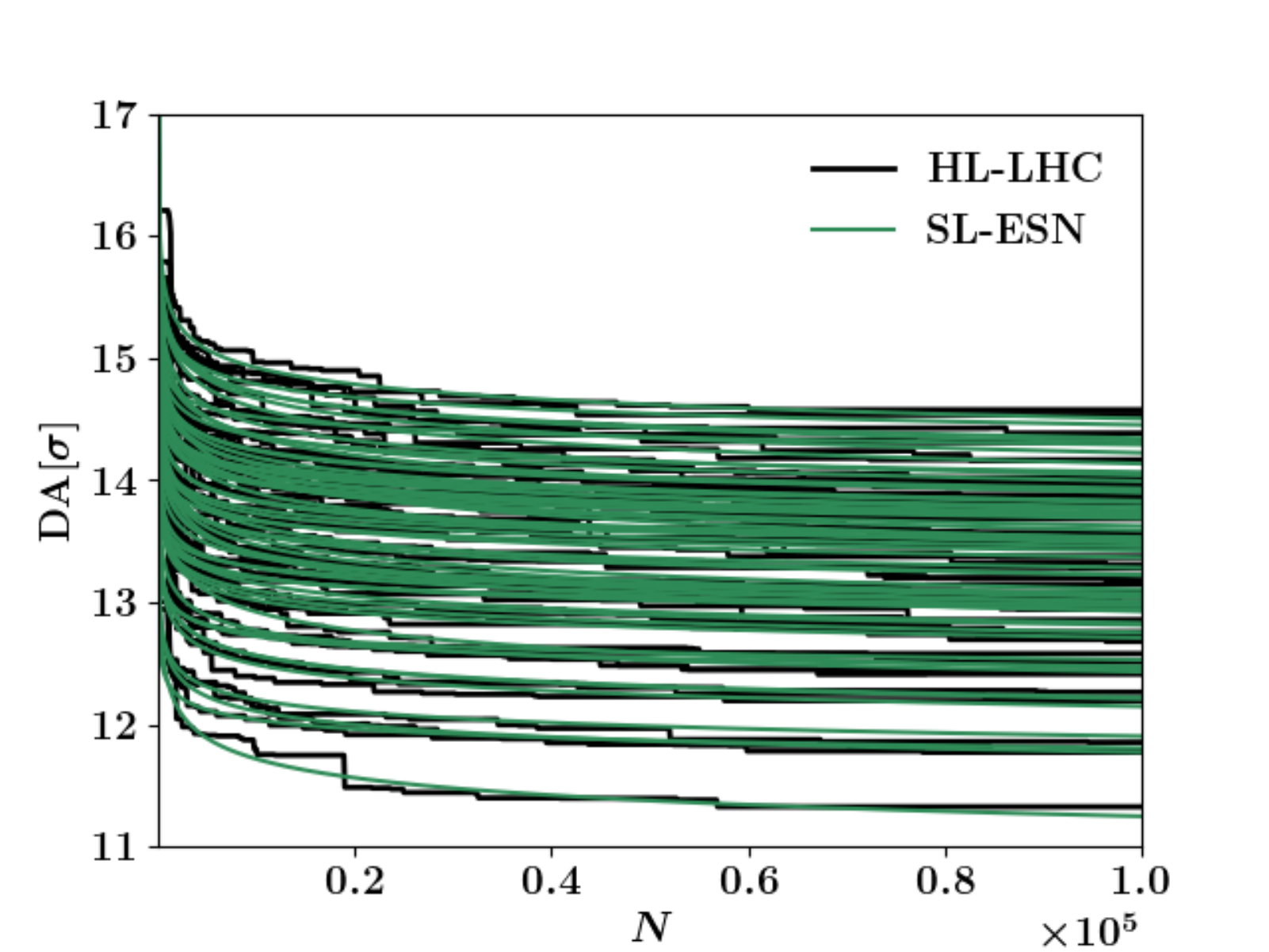}
\endminipage\hfill
\minipage{0.5\textwidth}
\centering
  \includegraphics[width=5.3cm]{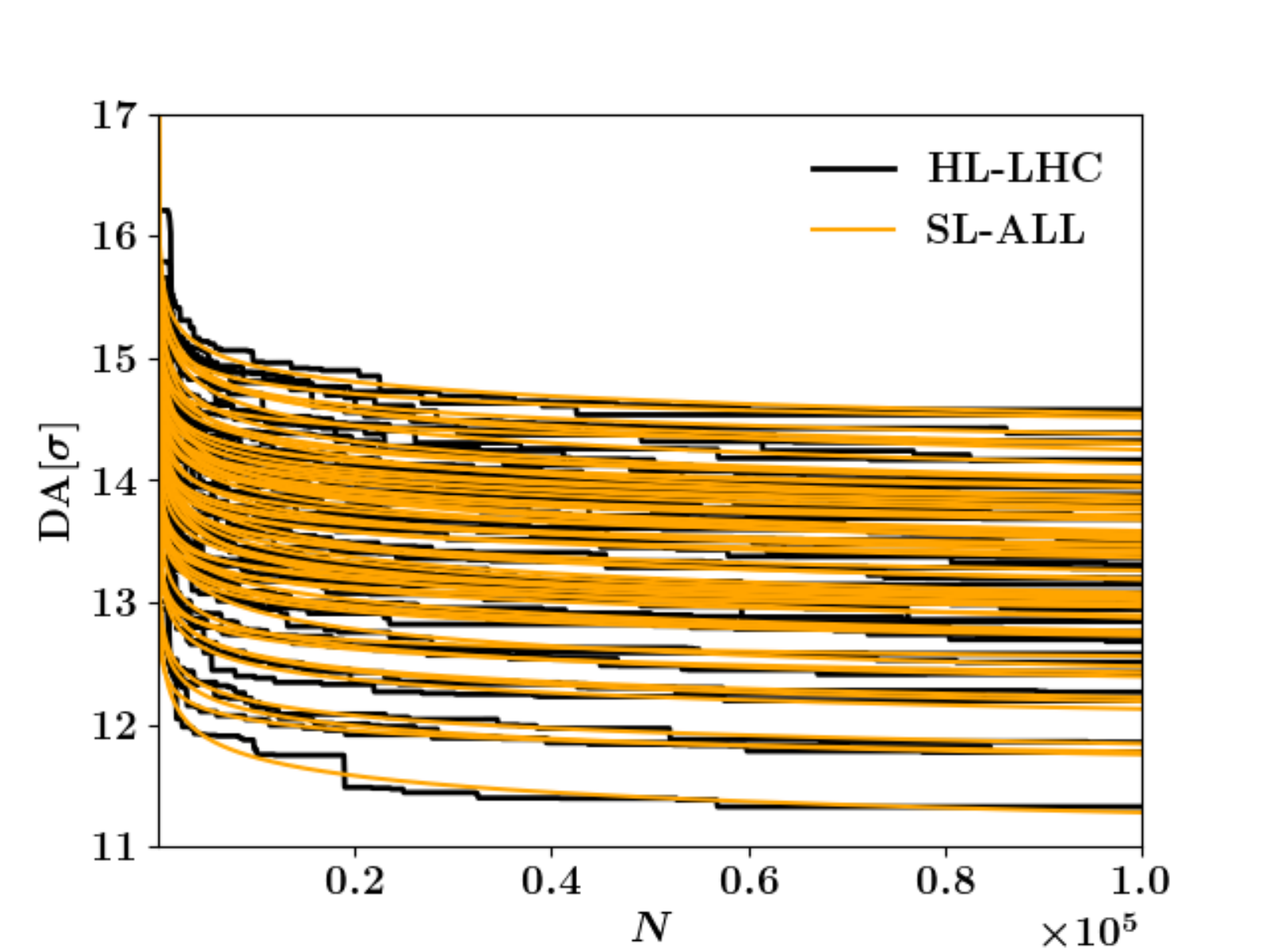}
\endminipage\hfill
\caption{Predictions for SL-ESN (left) and SL-ALL (right) for $N_\mathrm{d}$ = 60 seeds.}
\label{fig:hlslesn}
\end{figure}

\begin{figure}[!htb]
\centering
  \includegraphics[width=5.3cm]{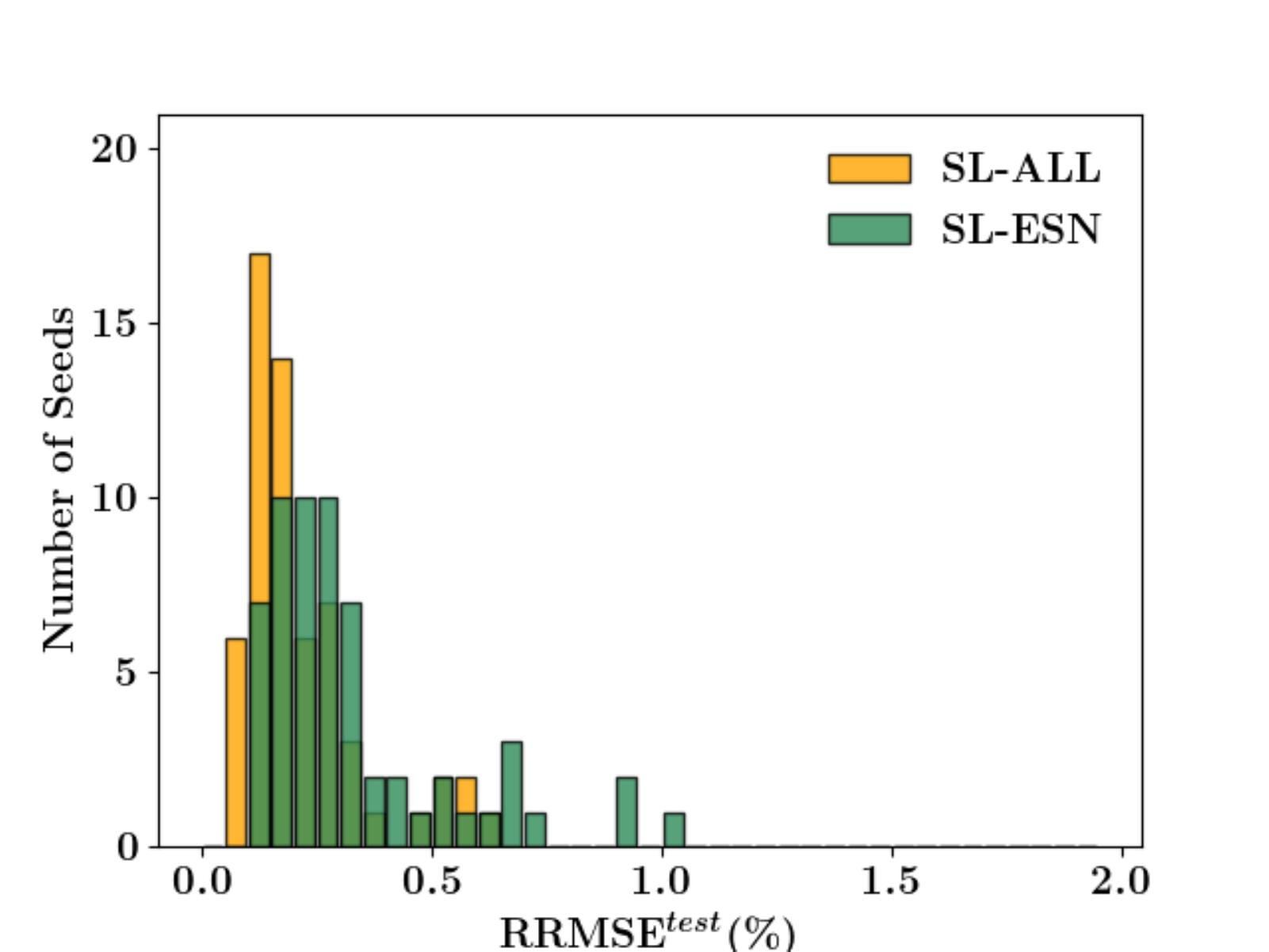}
\caption{Distribution of $\mathrm{RRMSE^{test}}$ for $N_\mathrm{d} = 60$ seeds for SL-ESN and SL-ALL.}
\label{fig:hlslesndist}
\end{figure}

In this section, we consider whether ESN predictions can possibly be used to replace the tracking simulations that generated the data in the \textit{test set}. In this sense, we fit the SL to the $k_\mathrm{fit}$ data plus the ESN predictions in the \textit{test set}. We denote this fit procedure by SL-ESN and compare it with the results of SL-ALL, which represents the best results that can be achieved with the SL approach\footnote{We recall that SL-ALL denotes the fit obtained by using all the available DA data, namely $k_\mathrm{fit}+k_\mathrm{test}$.}. The idea is to check the quality of the approximation of SL-ESN in the \textit{test set}, in view of further 
prediction beyond this set. The predictions provided by SL-ESN and SL-ALL for the $N_\mathrm{d}$ = 60 seeds can be seen in Fig.~\ref{fig:hlslesn} and the distribution of $\mathrm{RRMSE^{test}}$ is shown in Fig.~\ref{fig:hlslesndist}, while the mean, maximum, minimum, and standard deviation of $\mathrm{RRMSE^{test}}$ in Table~\ref{tab:hlesncompa}.

As it might be expected, all indicators of the distribution of $\mathrm{RRMSE^{test}}$ for SL-ESN are significantly larger than those for SL-ALL, as the first approach fits the prediction of ESN, not the real DA data. In fact, SL-ESN is essentially equivalent to ESN alone and hence more stable than SL alone as far as outliers are concerned. In other words, the SL-ESN seems to be an effective surrogate model that improves the predictions given by the SL only.

\begin{table}[htb]
\centering
\caption{Mean, maximum, minimum, and standard deviation of the $\mathrm{RRMSE^{test}}$ distribution.}
\begin{tabular}{|c|c|c|c|c|} 
\hline
 &  Mean & Max & Min & Std \\
\hline
\hline
SL-ESN & 0.33 & 1.01 & 0.10 & 0.21 \begin{tabular}[c]{@{}c@{}}\end{tabular} \\
\hline
SL-ALL & 0.21 & 0.64 & 0.06 &  0.13 \begin{tabular}[c]{@{}c@{}}\end{tabular} \\
\hline
\end{tabular}
\label{tab:hlesncompa}
\end{table}

\begin{figure}[!htb]
\minipage{0.5\textwidth}
\centering
  \includegraphics[width=5.3cm]{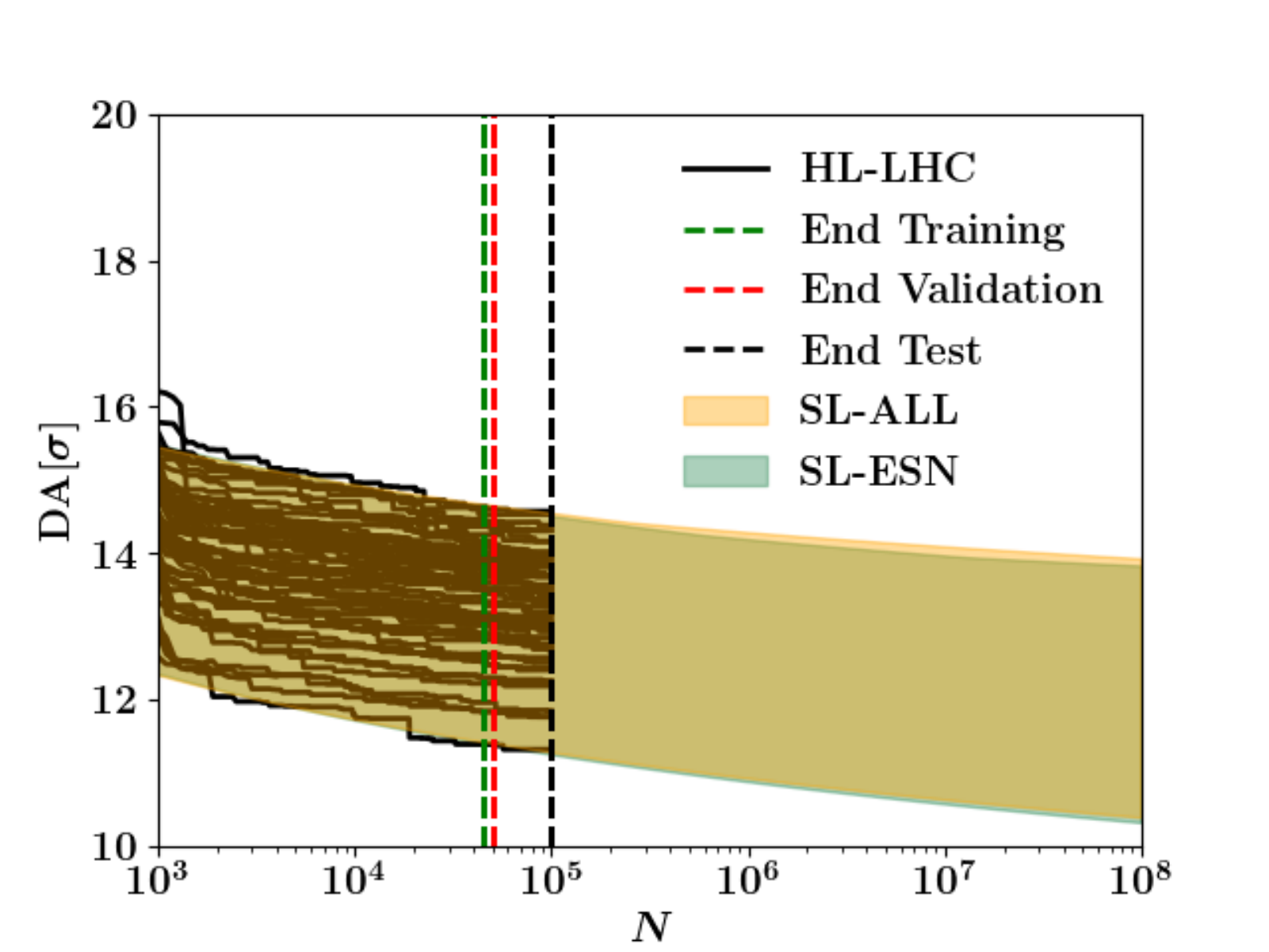}
\endminipage\hfill
\minipage{0.5\textwidth}
\centering
  \includegraphics[width=5.3cm]{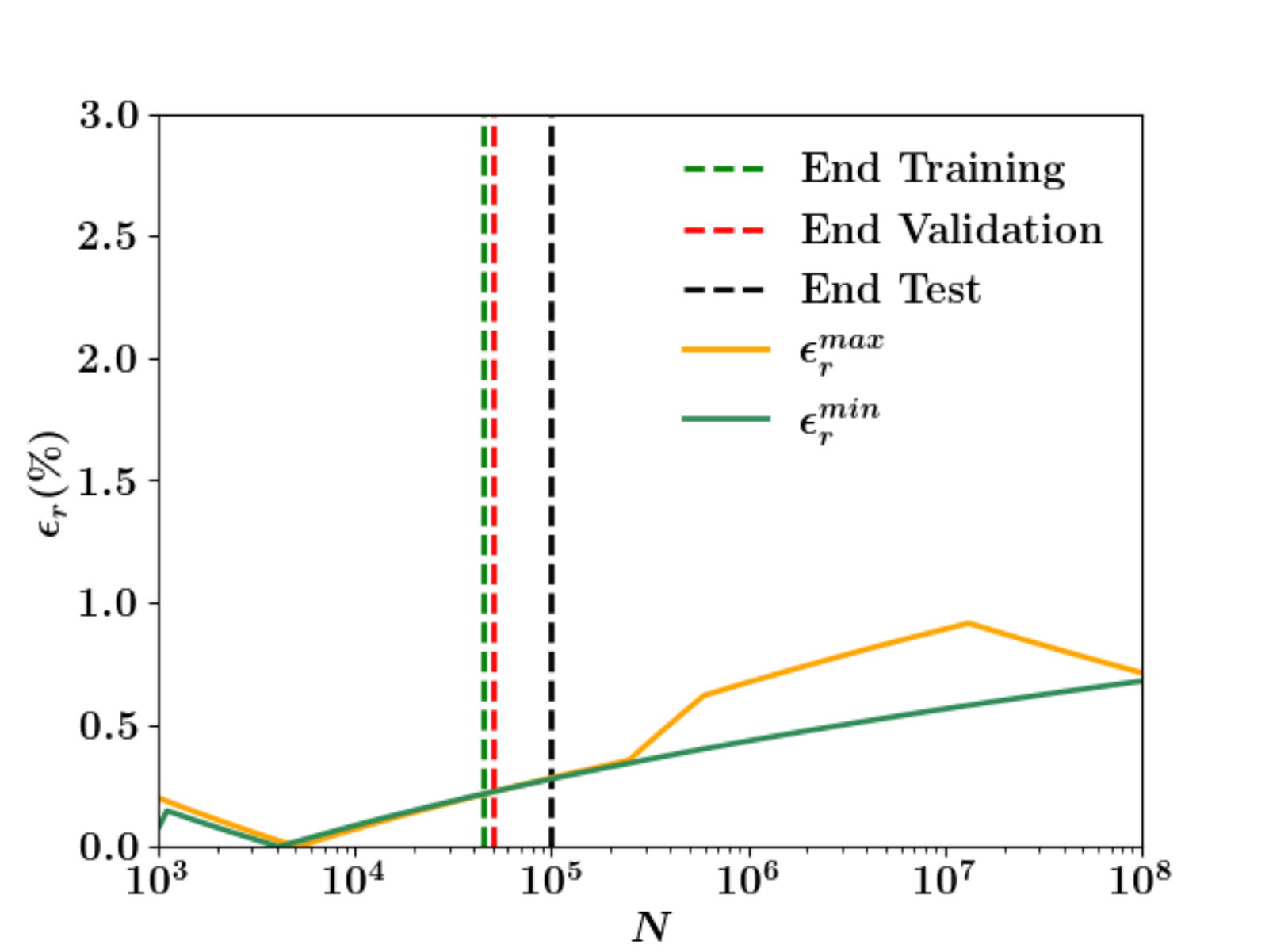}
\endminipage\hfill
\caption{Left: Envelope, i.e. minimum and maximum values of the SL-ESN and SL-ALL predictions extrapolated beyond the \textit{test set}. Right: Relative error $\epsilon_\mathrm{r}$ of the minimum and maximum DA predictions up to $N = 10^{8}$ turns.}
\label{fig:hllhclarge}
\end{figure}

After having evaluated the accuracy of the SL-ESN model in the \textit{test set}, we can check if it can replace the tracking simulations in this set. To do so, we compute predictions beyond the \textit{test set} and up to $N = 10^8$ turns. Since we do not have real DA data in this time interval, we cannot compute any metrics, and we use the envelope, i.e. minimum and maximum, of the predictions given by SL-ESN and SL-ALL to check whether SL-ESN approximates well the predictions given by SL-ALL beyond the \textit{test set}. We plot the envelope of the predictions given by SL-ESN and SL-ALL beyond the \textit{test set} in Fig.~\ref{fig:hllhclarge} (left), and we also show the relative error $\epsilon_\mathrm{r}$ defined as $\epsilon_\mathrm{r}^{i}$ = $\large( DA_\mathrm{SL-ALL}^{i}-DA_\mathrm{SL-ESN}^{i}\large)/DA_\mathrm{SL-ALL}^{i}$ where $i$ is $\mathrm{max}$ or $\mathrm{min}$ (right).

The two envelopes almost overlap until $N = 10^8$ turns, with $\epsilon_\mathrm{r}^\mathrm{max}$ and $\epsilon_\mathrm{r}^\mathrm{min}$ that are below $1\%$.  From this observation we conclude that we may only need to perform the tracking simulation until the end of the validation so that the tracking in the \textit{test set} could be spared. In fact, the predictions provided by SL-ESN are very similar to those of SL-ALL. In this way, we could use the ESN predictions to replace the tracking in the \textit{test set}. This result is in line with what was found in~\cite{giovannozzi:2021}, i.e. that the addition of synthetic points obtained by using Gaussian Processes improved the quality of the fitted SL model.

Running the SixTrack code~\cite{sixtrack,DeMaria:2711546} and the ESN model on the same CPU architecture, we have a speed-up of a factor $20$ by replacing the tracking simulations on $5\times 10~4$ turns, representing the \textit{test set}, with the prediction of the DA values by ESN. This evaluation of CPU time reduction can be easily improved by a trivial parallelisation of the ESN over the $100$ reservoirs.
Of course, the actual gain depends on several details, such as the model under consideration and the definition of the times that define the validation and test sets. It is worth stressing that whenever an actual accelerator lattice is used for the numerical DA computations, the CPU time needed depends not only on the number of turns used for the tracking, but also on the size of the accelerator, which corresponds approximately to the number of magnets comprised in the lattice, and on the characteristics of the magnetic field errors included in the accelerator model. In this respect, the computational gain implied by the proposed approach is even more relevant for the case of large future colliders, such as the Future Circular Hadron Collider (FCC-hh) under study at CERN~\cite{FCC-hhCDR,Benedikt:2022}. 

\subsection{DA Predictions for the H\'enon map data set}

To check the robustness of the current strategy, we apply it to a new system, which is the 4D H\'enon map introduced in Section~\ref{sec2}. 

\subsubsection{The ESN model}

Hyperparameters have been determined using the same approach as for the HL-LHC data and are reported in Table~\ref{tab:henonsens}. In this case, we also use $N_\mathrm{d} = 60$, but we have to stress that the various dynamics differ between them much more than the dynamics of the HL-LHC case. In fact, changes in the values of $\varepsilon$ and $\mu$ lead to radically different dynamical behaviours, whereas the HL-LHC realisations are much closer to each other, representing minor variations of the same dynamical behaviour. 

Only the values of $\Delta t$ and $\beta$ are different from those of the HL-LHC case.  Note that the value of $\beta$ found is much lower than that of HL-LHC. This means that the model is less overfitting than with the HL-LHC data, especially because the H\'enon DA data are much smoother. 
\begin{table}[htb]
\centering
\caption{Set $H$ of the hyperparameters tuned after validation using H\'enon map DA data.}
\begin{tabular}{|c|c|c|c|c|c|c|c|c|} 
\hline
$N_\mathrm{r}$ & $\beta$ & $\rho$ & $a$ & $BI$ & $L$ & $\Delta t$  & $f$ & $s$\\
\hline
\hline
20 & $9.10^{-6}$ & 0.99 & 1 & 0 & 1 & 0.004 & $\tanh$& 0 \begin{tabular}[c]{@{}c@{}}\end{tabular} \\
\hline
\end{tabular}
\label{tab:henonsens}
\end{table}

\begin{figure}[!htb]
\minipage{0.5\textwidth}
\centering
  \includegraphics[width=5.3cm]{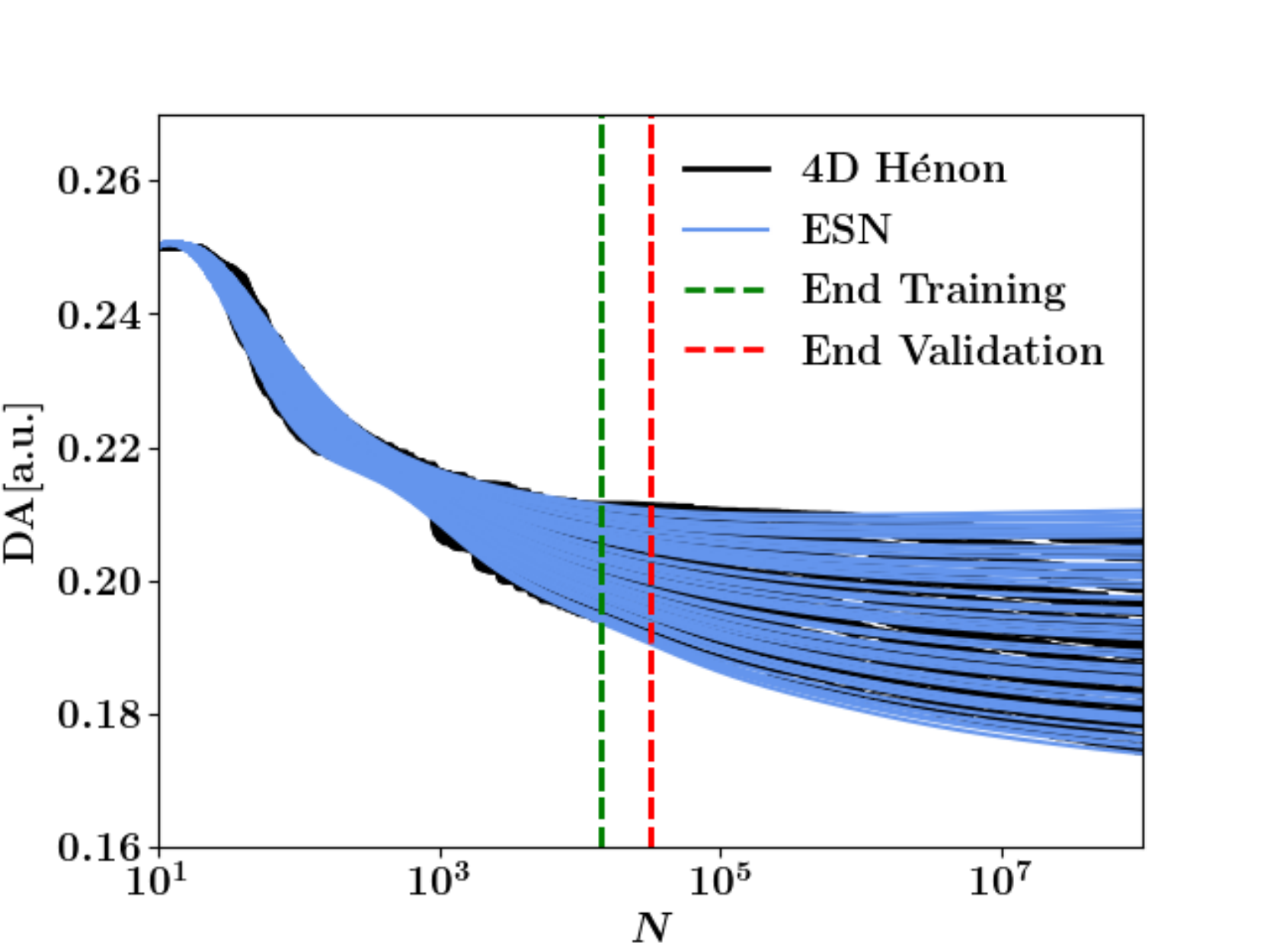}
\endminipage\hfill
\minipage{0.5\textwidth}
\centering
  \includegraphics[width=5.3cm]{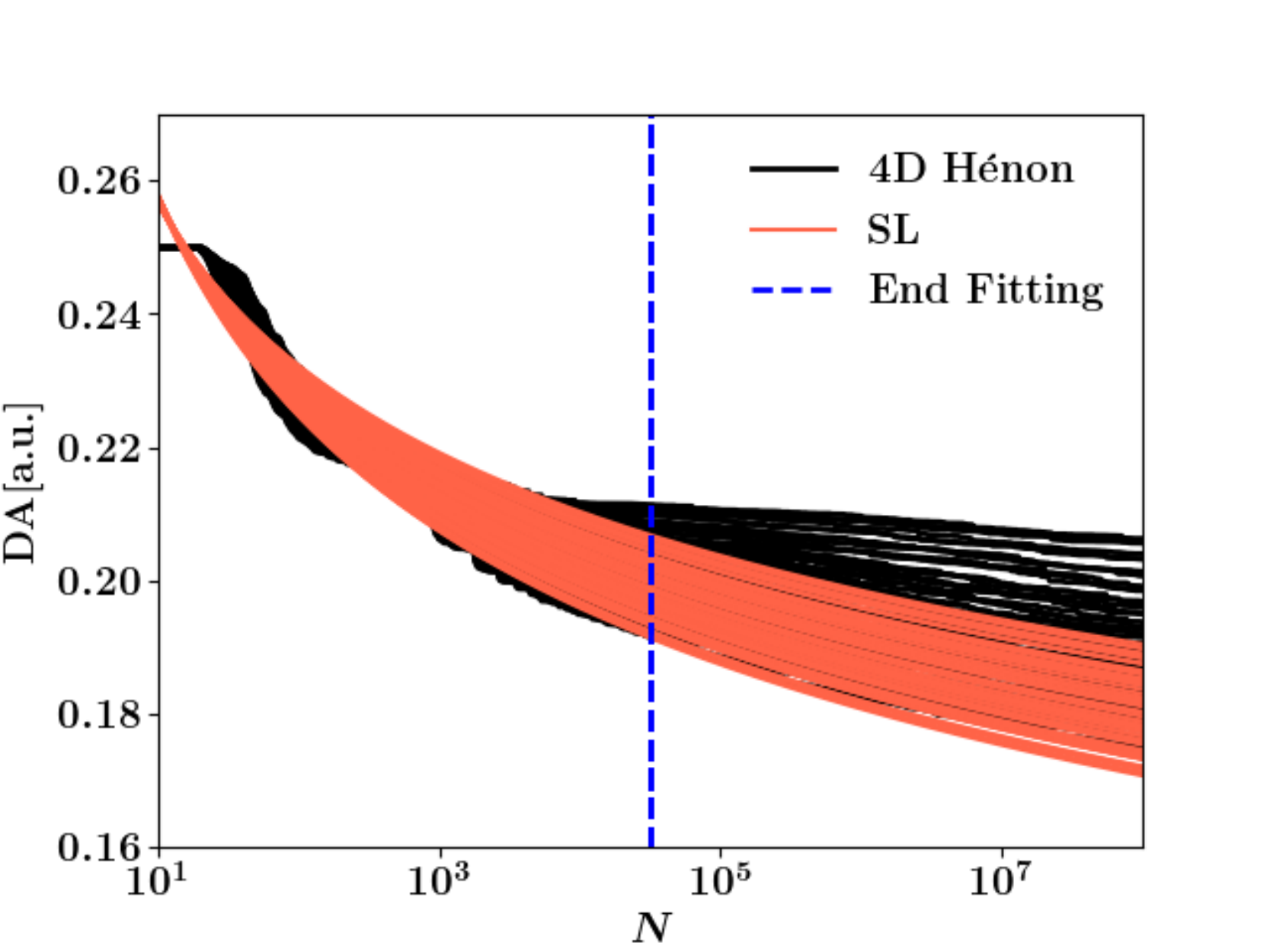}
\endminipage\hfill
\caption{DA predictions for ESN (left) and SL (right) for $N_\mathrm{d}$ = 60 seeds.}
\label{fig:henoncompa}
\end{figure}

In Fig.~\ref{fig:henoncompa}, we plot the $N_\mathrm{d} = 60$ DA predictions given by ESN and SL. For ESN, we recall that we used $k_{\mathrm{train}}=450$ and $k_{\mathrm{val}}=50$ data, and for SL we used the $k_{\mathrm{fit}}=500$ data. Furthermore, \textit{test set} is the same for both ESN and SL. As we can see, the SL predictions in the \textit{test set} do not perform well, whereas those provided by the ESN fit the training/validation/test data much better. 

\begin{figure}[!htb]
\centering
  \includegraphics[width=5.3cm]{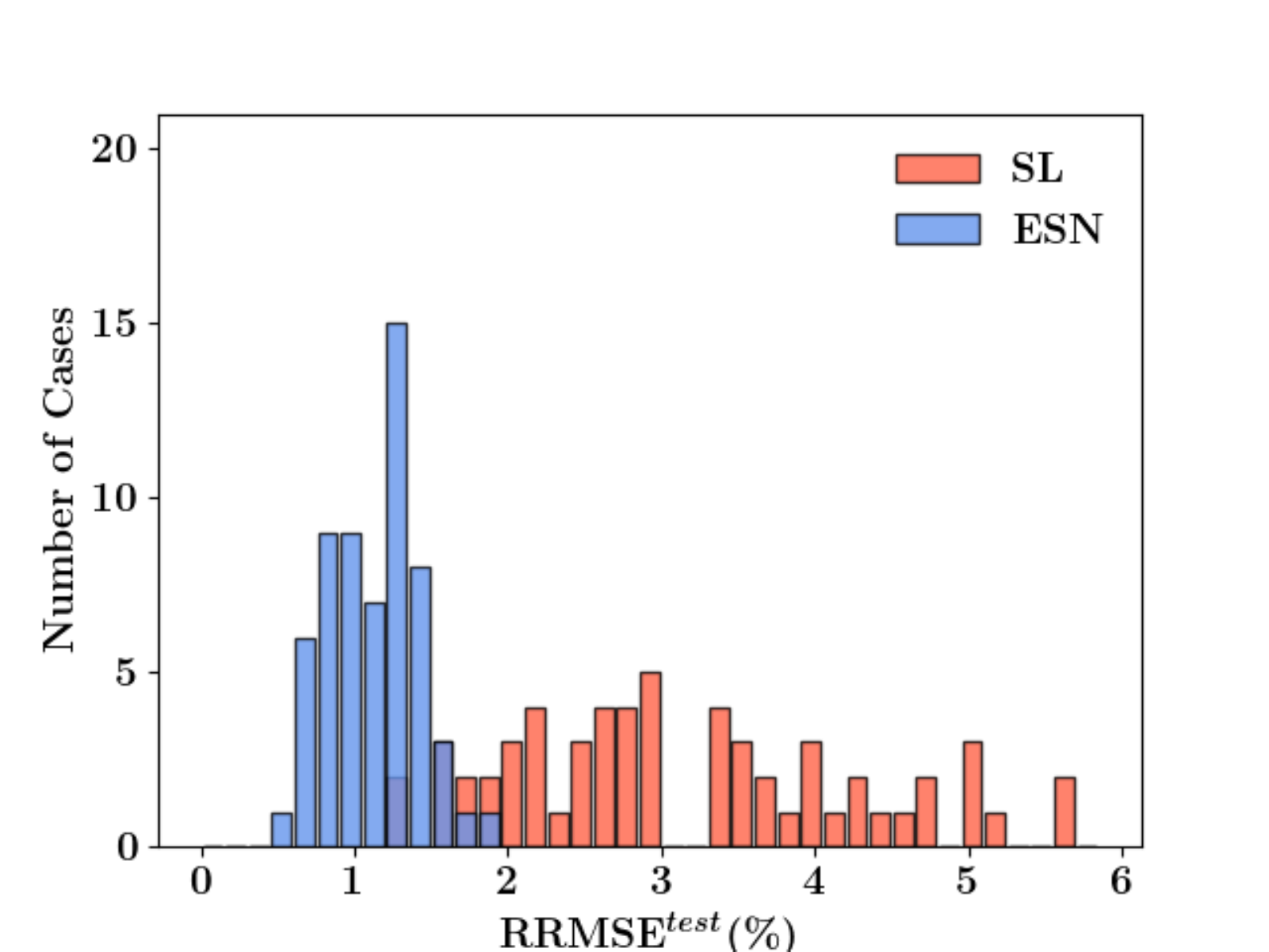}
\caption{Distribution of $\mathrm{RRMSE^{test}}$ for $N_\mathrm{d}$=60 seeds for ESN and SL.}
\label{fig:henondist}
\end{figure}

In Fig.~\ref{fig:henondist} we compare the distributions of $\mathrm{RRMSE^{test}}$ for ESN and SL, and the first is clearly much narrower and closer to zero than the latter. This behaviour is easily explained by considering the fact that the scaling law is an asymptotic law that aims to describe the long-term behaviour of the DA (using very few model parameters). Therefore, it is not effective in reproducing the detailed behaviour of the DA for low numbers of turns. Our ESN model is able to fit both the short-term and long-term behaviour simultaneously, thus explaining the observed better performance.  

The mean, maximum, minimum, and standard deviation of $\mathrm{RRMSE^{test}}$ for the two approaches are reported in Table~\ref{tab:henoncompa}. 

\begin{table}[htb]
\centering
\caption{Mean, maximum, minimum, and standard deviation of the $\mathrm{RRMSE^{test}}$ distribution.}
\begin{tabular}{|c|c|c|c|c|c|c|c|c|} 
\hline
 &  Mean & Max & Min & Std \\
\hline
\hline
ESN & 1.13 & 1.89 & 0.59 & 0.28 \begin{tabular}[c]{@{}c@{}}\end{tabular} \\
\hline
SL & 3.17 & 5.85 & 1.25 &  1.18 \begin{tabular}[c]{@{}c@{}}\end{tabular} \\
\hline
\end{tabular}
\label{tab:henoncompa}
\end{table}
The table shows, in a quantitative way, the differences observed in the histogram of the distributions. In fact, the RRMSE of the ESN is on average about 3 times lower than that of the SL, which is a significant improvement compared to the case of HL-LHC. Several reasons can explain this behaviour. First, the DA data for the H\'enon map are much smoother than those of the HL-LHC data set, which improves training and limits overfitting of the ESN. Second, as already mentioned, the behaviour of the $N_\mathrm{d}$ dynamics is very diverse, and the SL, with only two free parameters, is clearly disadvantaged with respect to the ESN. Moreover, since the SL is an asymptotic law, its performance has been downgraded by including low-turn DA data.

\subsubsection{The SL-ESN model}

\begin{figure}[!htb]
\minipage{0.5\textwidth}
\centering
  \includegraphics[width=5.3cm]{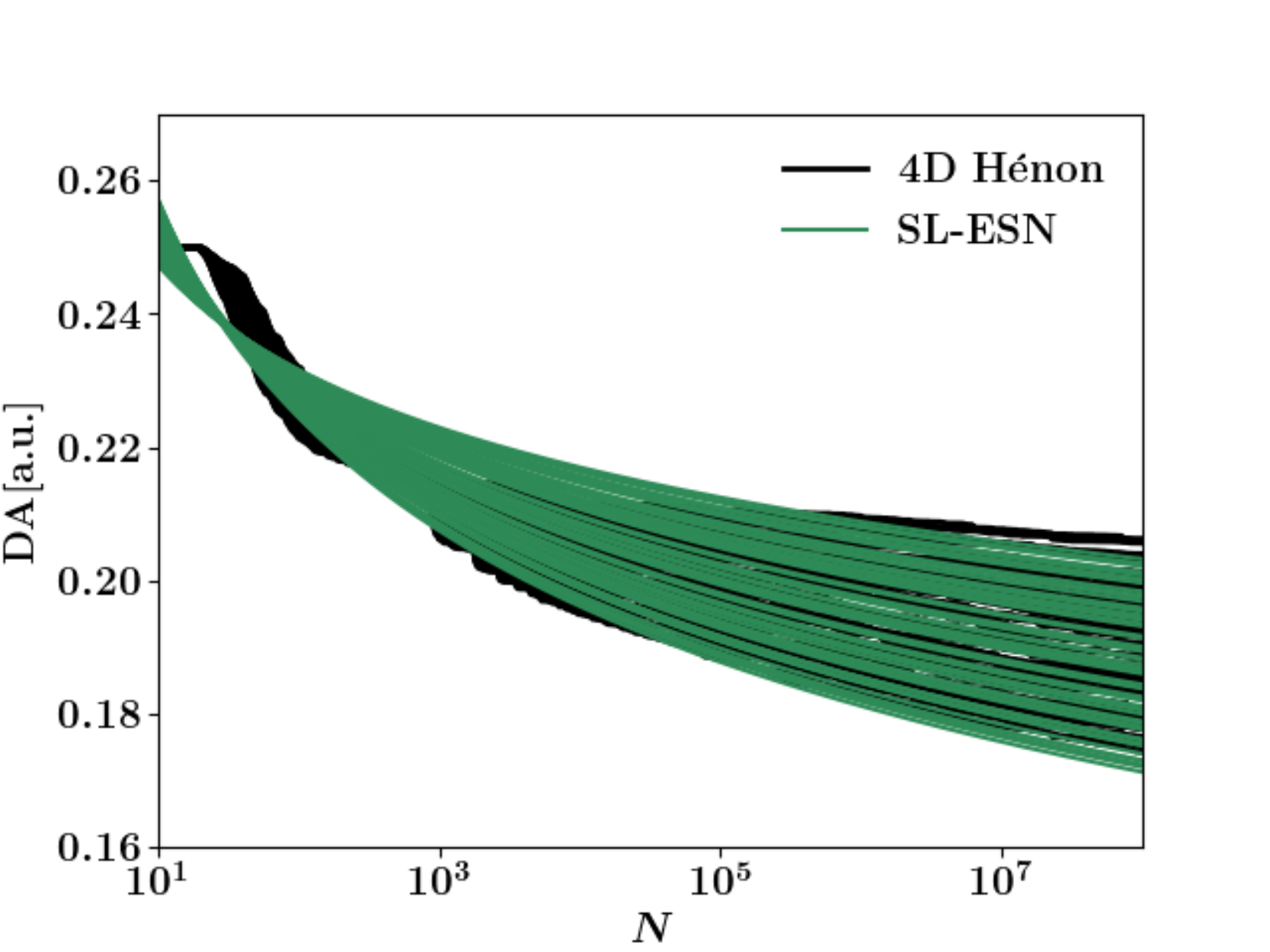}
\endminipage\hfill
\minipage{0.5\textwidth}
\centering
  \includegraphics[width=5.3cm]{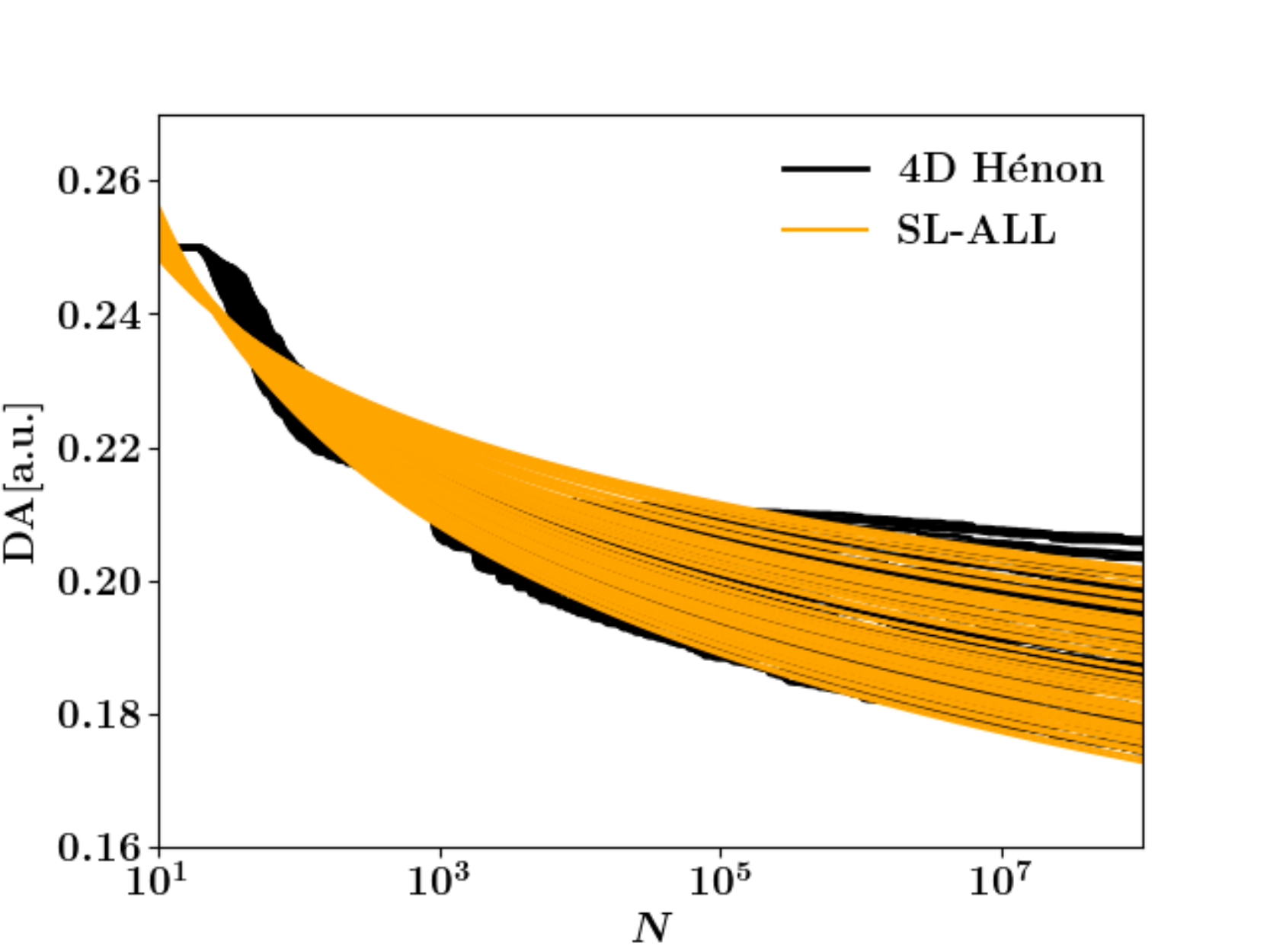}
\endminipage\hfill
\caption{Predictions for SL-ESN (left) and SL-ALL (right) for $N_\mathrm{d}$ = 60 seeds.}
\label{fig:henonslesn}
\end{figure}

\begin{figure}[!htb]
\centering
  \includegraphics[width=5.3cm]{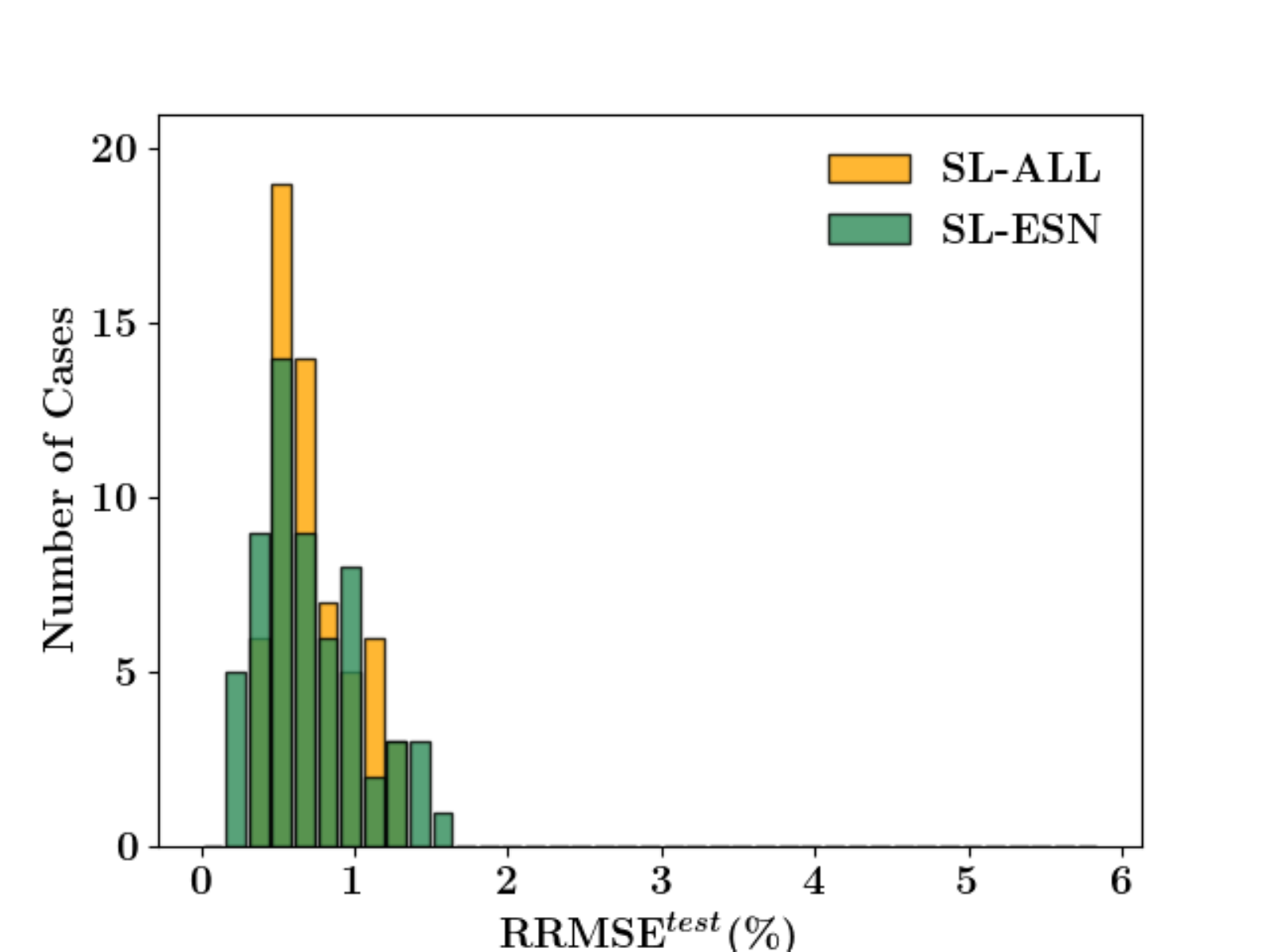}
\caption{Distribution of $\mathrm{RRMSE^{test}}$ for $N_\mathrm{d} = 60$ seeds for SL-ESN and SL-ALL.}
\label{fig:henonslesndist}
\end{figure}

We repeat the procedure to check if the ESN predictions can replace the tracking simulation in the \textit{test set}. As previously, we compare SL-ESN with SL-ALL. The predictions given by SL-ESN and SL-ALL for the 60 cases can be seen in Fig.~\ref{fig:henonslesn}, the distribution of $\mathrm{RRMSE^{test}}$ is shown in Fig.~\ref{fig:henonslesndist}, and the mean, maximum, minimum, and standard deviation of $\mathrm{RRMSE^{test}}$ are reported in Table~\ref{tab:henonallcompa}.
\begin{table}[htb]
\centering
\caption{Mean, maximum, minimum, and standard deviation of the $\mathrm{RRMSE^{test}}$ distribution.}
\begin{tabular}{|c|c|c|c|c|c|c|c|c|} 
\hline
 &  Mean & Max & Min & Std \\
\hline
\hline
SL-ESN & 0.71 & 1.57 & 0.26 & 0.33 \begin{tabular}[c]{@{}c@{}}\end{tabular} \\
\hline
SL-ALL & 0.72 & 1.29 & 0.35 &  0.24 \begin{tabular}[c]{@{}c@{}}\end{tabular} \\
\hline
\end{tabular}
\label{tab:henonallcompa}
\end{table}

In this case, the SL-ESN performs equally well as the SL-ALL. In fact, the mean of $\mathrm{RRMSE^{test}}$ is the same. Furthermore, fitting the SL to the predictions of the ESN allows us to improve the accuracy of the ESN and the SL. Taking into account the average, SL-ESN is almost 2 times and 4 times more accurate than ESN and SL, respectively. Similarly to the HL-LHC case, the standard deviation and maximum $\mathrm{RRMSE^{test}}$ of SL-ESN are much lower than those of SL, which shows a certain robustness of the conclusions that SL-ESN helps improve SL.

\begin{figure}[!htb]
\minipage{0.5\textwidth}
\centering
  \includegraphics[width=5.3cm]{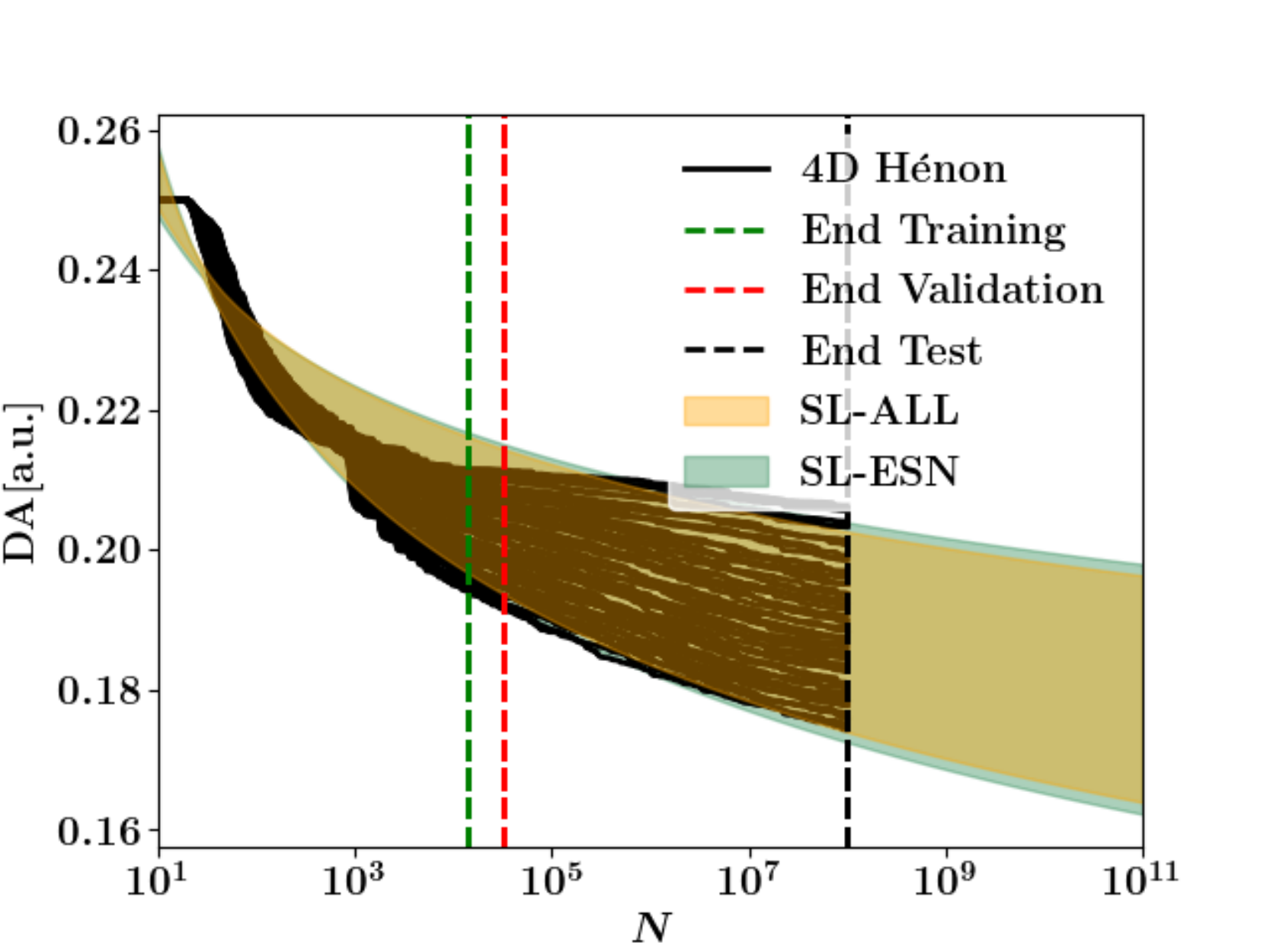}
\endminipage\hfill
\minipage{0.5\textwidth}
\centering
  \includegraphics[width=5.3cm]{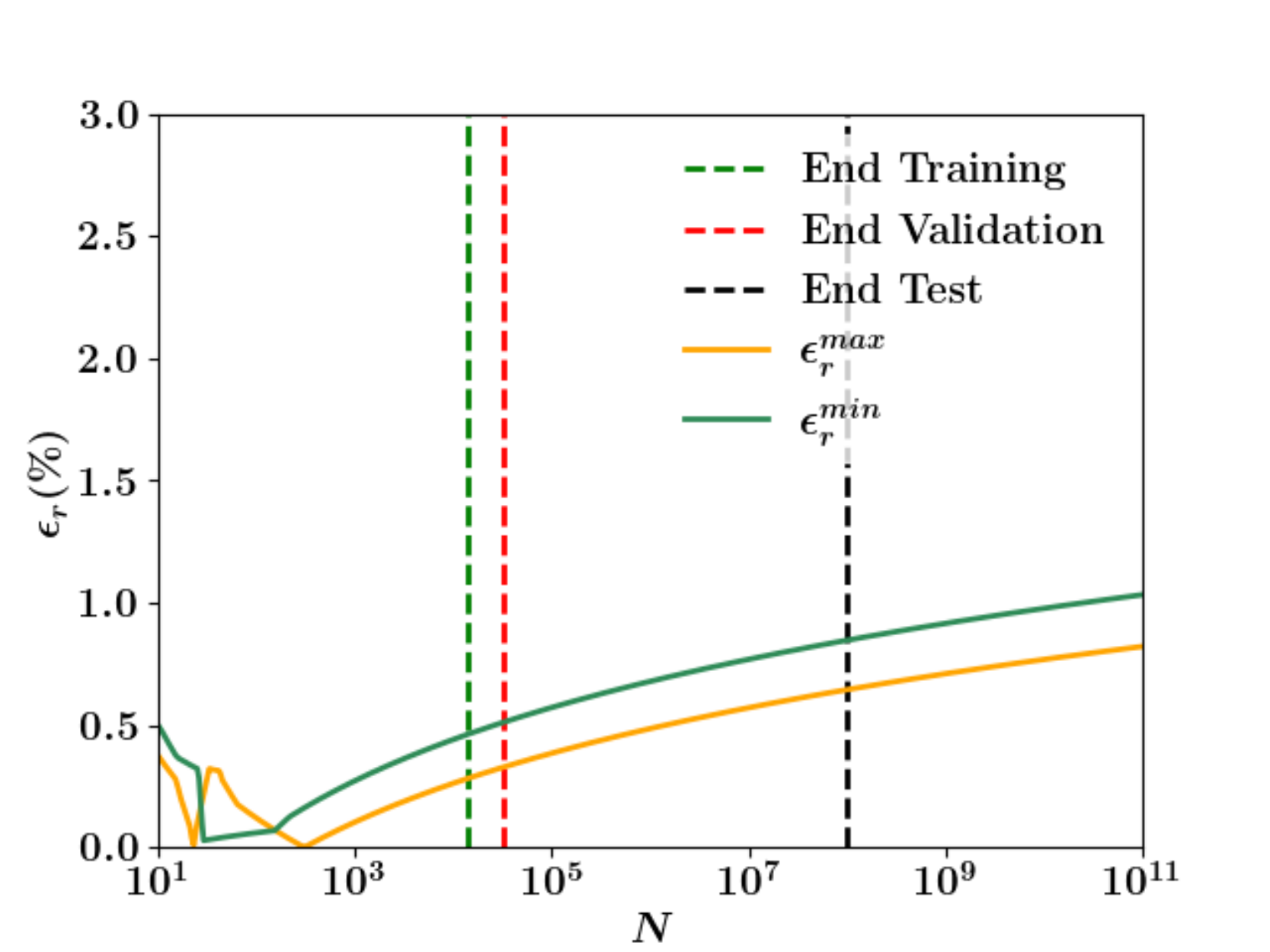}
\endminipage\hfill
\caption{Left: Envelope, i.e. minimum and maximum values of the SL-ESN and SL-ALL predictions extrapolated beyond the \textit{test set}. Right: Relative error $\epsilon_\mathrm{r}$ of the minimum and maximum DA predictions up to $N = 10^{11}$ turns.}
\label{fig:henonlarge}
\end{figure}

To further check whether the ESN predictions can replace the tracking simulation in the \textit{test set}, we perform the prediction beyond the \textit{test set} up to $N = 10^{11}$ turns. As previously, we do not have the real DA data in this range, so we cannot compute any metrics. We plot the envelope of the predictions given by SL-ESN and SL-ALL in Fig.~\ref{fig:henonlarge}.

The two envelopes of the predictions almost overlap until $N = 10^{11}$, and the relative errors $\epsilon_\mathrm{r}^\mathrm{max}$ and $\epsilon_\mathrm{r}^\mathrm{min}$ are below $1.5\%$, as for the case HL-LHC. This indicates, once again, that the tracking simulation in the \textit{test set} could be replaced by the ESN predictions. 

\section{Conclusions}\label{conc}

In this article, we have presented the results obtained with an ensemble approach to ESN reservoir computing for the prediction of the dynamic aperture of a circular hadron accelerator. In particular, we have compared the performance of ESN with that of a scaling law based on the Nekhoroshev theorem to predict the evolution of the dynamic aperture over time. This analysis has been carried out on two data sets that have been generated using numerical simulations performed on realistic models of the transverse beam dynamics in the HL-LHC and on a modulated 4D H\'enon map with quadratic and cubic non-linearities. 

We have shown that the average accuracy in the \textit{test set} of the scaling law used to fit the ESN predictions was better than that of the scaling law alone. In particular, we have observed that the standard deviation of the RRMSE of the scaling law combined with the ESN is much lower than that of the scaling law alone. This leads to more reliable predictions. The fact that this observation is confirmed for both data sets gives us confidence that the combination of the scaling law and the ESN is the best approach. 

A consequence of this result is that the tracking performed in the \textit{test set} can be avoided by replacing it with the predictions of the ESN. In fact, for both the HL-LHC and H\'enon map data sets, the predictions of the scaling law combined with the ESN and of the scaling law fitted to the entire data set are close to the percent level, even for numbers of turns three orders of magnitude beyond that of the \textit{test set}. The gain in CPU time depends on the size of the accelerator and the complexity of its model. However, it is clear that the proposed approach is particularly appealing for hadron colliders of the post-LHC era that are currently being studied. 

The study presented here represents only the beginning of a research area that could be further developed in the future given the promising results obtained. The partition of available data into training, validation, and test data sets should be studied in more detail to assess whether such a partition could be obtained using an appropriate algorithm. The established link between dynamic aperture and models for the evolution of intensity in hadron rings and the evolution of luminosity in hadron colliders could be further developed by using the promising results discussed in this paper. Investigations on the possibility of using ESN to improve the modelling of beam lifetime and luminosity evolution should be seriously considered and pursued. Finally, the predictive power of ESN could be applied to indicators of chaos, which are dynamical observables computed over the orbit of an initial condition to establish whether the motion is regular or chaotic, to improve their performance. This would be another important topic that could bring important insight to the field of non-linear beam dynamics.

\section*{Acknowledgements}
We would like to express our gratitude to C.E.~Montanari for providing the software to perform dynamic aperture simulations for the 4D H\'enon map.

\newpage
\begin{appendices}

\section{The Echo State Property}\label{AP:ESP}

An important prerequisite for the output-only training is the so-called Echo State Property (ESP), which guarantees that initial conditions have an effect that vanishes over time. We use the results presented in~\cite{ESP} to recall the definition of ESP and a new sufficient condition that can be used in practice. In fact, satisfying the ESP allows us to guarantee that the reservoir activation state $x_{k-1}$ is uniquely determined by any left-infinite input sequence $\ldots, u_{k-2},u_{k-1}$.

To define ESP, we require the \textit{compactness condition}, that is, we assume that the states and inputs belong to compact sets $\mathcal{X} \subset \mathbb{R}^{N_\mathrm{r}}, \mathcal{U} \subset \mathbb{R}^{K}$ and that 
 $F(x_{k-1},u_k) \in \mathcal{X}$ and $u_{k-1} \in \mathcal{U}, \, \forall k \in \mathbb{Z} $.
In practice, the ESN inputs will always be bounded, so the compactness of $\mathcal{U}$ is guaranteed. Furthermore, for bounded sigmoid functions $f$, such as $\tanh$, the state space $\mathcal{X}$ is also compact.
We define $ \mathcal{U}^{-\infty} = \{ u^{-\infty} = (\ldots , u_{-1},u_0), \, u_{k} \in \mathcal{U} \, \forall k \in \mathbb{Z} \} $ and $ \mathcal{X}^{-\infty} = \{x^{-\infty} = (\ldots, x_{-1},x_0), \, x_k \in X \, \forall k \in \mathbb{Z}\}$, which are the sets of infinite left input and reservoir activation state sequences.

\textbf{Definition 3.1}\label{def:ESP} (ESP). A network $F : \mathcal{X} \times \mathcal{U} \rightarrow \mathcal{X}$ with the \textit{compactness condition} has the ESP with respect to $\mathcal{U}$ if for any left input sequence $u^{-\infty} \in \mathcal{U}^{-\infty}$ and any two state sequences $x^{-\infty}, y^{-\infty} \in \mathcal{X}^{-\infty}$ compatible with $u^{-\infty}$ (i.e. $x_k = F(x_{k-1},u_k), \forall k \leq 0$), then for all $k \geq 0, \| x_k - y_k\| \leq \delta_k$, where $\delta_k$ denotes a small value.

Definition~\ref{def:ESP} is not easily applicable in practice. Thus, we introduce the following Theorem~\ref{theoreme:Sufficient ESP} that should be used in practise as it provides a sufficient condition to satisfy the ESP in the case of a leaky ESN: 

\textbf{Theorem 3.1}\label{theoreme:Sufficient ESP} (Sufficient condition of the ESP). If the spectral radius of the matrix $$\Tilde{W} = \frac{\Delta t}{c}W+\left (1 -a\frac{\Delta t}{c}\right)I$$ is smaller than 1, then the leaky ESN with $f$ = $\tanh$ satisfies the ESP for all inputs. However, this condition is only sufficient, but not necessary. In other words, setting $\rho(\Tilde{W}) \geq 1$ does not necessarily lead to poor performance of leaky ESN.

\end{appendices}

\newpage

\bibliographystyle{bst/sn-mathphys}
\bibliography{sn-bibliography}

\end{document}